\documentclass[preprintnumbers,twocolumn]{revtex4}
\usepackage{dcolumn}
\usepackage{bm}
\usepackage{amsfonts}
\usepackage{amsmath}
\usepackage{amssymb}
\usepackage{subfigure}
\usepackage{graphicx}
\usepackage[dvipdfm]{epsfig}

\begin{document}

\title{Quantum theory of Synchronously Pumped type I Optical Parametric
Oscillators: characterization of the squeezed supermodes}
\author{Giuseppe Patera$^{1}$, Nicolas Treps$^{1}$, Claude Fabre$^{1}$, Germ%
\'{a}n J. de Valc\'{a}rcel$^{1,2}$}
\affiliation{$^{1}$Laboratoire Kastler Brossel, Universit\'{e} Pierre et Marie
Curie-Paris6, ENS, CNRS, 4 place Jussieu CC74, 75252 Paris cedex 05, France}
\affiliation{$^{2}$Departament d'\`{O}ptica, Universitat de Val\`{e}ncia, Dr. Moliner 50,
46100 Burjassot, Spain}

\begin{abstract}
Quantum models for synchronously pumped type I optical parametric
oscillators (SPOPO) are presented. The study of the dynamics of SPOPOs,
which typically involves millions of coupled signal longitudinal modes, is
significantly simplified when one considers the ``supermodes", which are
independent linear superpositions of all the signal modes diagonalizing the
parametric interaction. In terms of these supermodes the SPOPO dynamics
becomes that of about a hundred of independent, single mode degenerate OPOs,
each of them being a squeezer. One derives a general expression for the
squeezing spectrum measured in a balanced homodyne detection experiment,
valid for any temporal shape of the local oscillator. Realistic cases are
then studied using both analytical and numerical methods: the oscillation
threshold is derived, and the spectral and temporal shapes of the squeezed
supermodes are characterized.
\end{abstract}

\maketitle

\section{Introduction}

%________________________________________

Mode-locked trains of pulses, or frequency combs, have at the same time the
coherence properties of c.w. lasers and the high peak powers of pulsed
lasers. They are potentially perfect tools for generating non-classical
states of light, as they are are at the same time high quality light
sources, free of excess noise, and intense sources able to induce strong
nonlinear effects, and therefore to generate strongly non-classical states
of light, such as squeezed or quadrature-entangled states.

Frequency combs have been used in many quantum optics experiments, and have
efficiently produced non classical light, either in $\chi ^{(2)}$ \cite%
{Grangier,ManipSPOPO6,Serkland} or $\chi ^{(3)}$ \cite%
{Watanabe,Rosenbluh,Friberg,Schmitt,Krylov,Hirosawa,Sharping,Fiorentino} media,
but so far in a single-pass configuration in the nonlinear medium. In this
configuration, one needs very high peak powers, and the system loses somehow
its potential high quality in terms of pulse to pulse coherence and
transverse profile. Mode locked lasers have also been used to efficiently
generate squeezed states in optical fibers \cite{Levenson,Shelby}. The
system has the drawback of generating non-minimal states states of light,
because of the excess noise due to Brillouin scattering in the fiber.

We have recently proposed \cite{ourSPOPO06} to use synchronous optical
cavities to recirculate the light in the nonlinear medium, thus enhancing
further the nonlinear effects and imposing the cavity mode structure to the
generated non-classical field. This can be done by building ``synchronously
pumped OPOs" or SPOPOs. In a SPOPO the cavity round-trip time is equal to
the delay between successive pulses of the pumping mode-locked laser, so
that the effect of the successive intense pump pulses add coherently, thus
reducing considerably its oscillation threshold. In \cite{ourSPOPO06}, a
large squeezing effect was predicted in some ``supermodes", which are well
defined linear combinations of signal modes of different frequency, but not
studied in detail. In the time domain these supermodes correspond to trains
of pulses of different waveforms, orthogonal each other. The purpose of this
paper is to precise the quantum model used to predict the effects and to
investigate in a detailed way, through analytical or numerical methods, the
potentialities of the system in realistic situations

SPOPOs have already been implemented as efficient sources of tunable
ultra-short pulses \cite%
{manipSPOPO0,manipSPOPO1,manipSPOPO2,manipSPOPO3,manipSPOPO4,manipSPOPO5}
their temporal properties have been theoretically investigated \cite%
{theorySPOPO1,theorySPOPO3,theorySPOPO4}, and actively mode-locking of OPOs
has been recently achieved \cite{Fabien}.

%Let us also mention that similar supermodes have been introduced in a different context \cite{BB,Polish},
%and that they have strong connections with the Schmidt and Bloch-Messiah decomposition [GIUSEPPE PRECISE],
%while frequency combs have been also proposed to generate cluster states\cite{pfister}.

The decomposition of a pulsed field in terms of a basis of normal
modes, similar to the supermodes we consider in this paper, has been
introduced in different contexts for a complete quantum characterization of
either the pulsed squeezed light generated by parametric down conversion
\cite{BB,Polish} or solitons in optical fibers \cite{Opatrny2002}. Such
approaches are strongly connected with the Schmidt decomposition of
two-photon states for the characterization of pairwise entanglement \cite%
{Huang1996,Law2000} and the Bloch-Messiah reduction of any optical circuit
characterized by a linear input-output relation \cite{Braunstein2005}.
In this context Menicucci \textsl{et al.} \cite{Menicucci2008}
proposed optical frequency combs as scalable resources for quantum
computation.

The article is organized as follows: we present first the model that we will
use. The system turns out to be characterized by a real and symmetric matrix
$\mathcal{L}$, which contains all the information about the effective
nonlinear interaction. The eigensystem of $\mathcal{L}$ is thus of special
relevance and is studied in Section III, where the SPOPO threshold and
several general properties of the spectrum of $\mathcal{L}$ are addressed.
An analytical approximation to the diagonalization is also given that allows
a better insight into the general trends as parameters are varied. In
Section IV it is shown that the introduced eigenmodes or supermodes are
squeezed, the corresponding eigenvalues determining the amount of squeezing,
which can be measured in a balanced homodyne detection experiment that uses
as the local oscillator (LO) a field with the same spectrum as the desired
supermode. In Section V one then studies the squeezing properties of SPOPOs in
two realistic cases, corresponding to BIBO and KNbO$_{3}$ crystals, using an
appropriate scaling property of the diagonalization problem. Finally,
an appendix details the case of the singly resonant SPOPO.

\section{The SPOPO model}

\subsection{Evolution equations for the operators}

We consider quasi-degenerate collinear type I interaction, by means of which
the pumping frequency comb, at frequencies around $2\omega _{0}$, is
converted by a nonlinear $\chi ^{\left( 2\right) }$ crystal into multimode
signal radiation at frequencies around $\omega _{0}$, and vice-versa, where $%
2\omega _{0}$ and $\omega _{0}$ are the two frequencies at which perfect
phase matching occurs. This implies that one has $n\left( 2\omega
_{0}\right) =n\left( \omega _{0}\right) \equiv n_{0}$, where $n$ is the
crystal refractive index. The nonlinear crystal is placed inside a high
finesse optical cavity of length $L$, which is assumed to be dispersion
compensated by intracavity dispersive elements, so that all cavity modes
around the frequency $\omega _{0}$ are equally spaced by a common free
spectral range $\Omega $, which is made equal to that of the pumping laser,
thus warranting the synchronization of the pump to the OPO cavity. This
ensures that the pulse-to-pulse delay of the pump beam coincides with the
cavity round-trip time and successive pump and signal pulses superpose in
time, thus maximizing the strength of the interaction. Hence the external
pump mean field, which is a phase-locked multimode coherent field, can be
written as
\begin{equation}
E_{\mathrm{ext}}\left( t\right) =\sqrt{\frac{P}{2\varepsilon _{0}c}}%
\sum_{m}i\alpha _{m}e^{-i\left( 2\omega _{0}+m\Omega \right) t}+\mathrm{c.c.}%
,  \label{Eext}
\end{equation}%
$P$ is the average laser irradiance (power per unit area), $\alpha _{m}$ is
the normalized ($\sum\nolimits_{m}\left\vert \alpha _{m}\right\vert ^{2}=1$)
complex spectral component of longitudinal mode labeled by the integer index
$m$, and $m=0$ corresponds to the phase-matched mode. As we will be
concerned with femtosecond lasers with pulse durations around $100$fs
, the number of pump modes will be typically on the order of $10^{4}-10^{5}$.

Two possibilities for pumping can be used: either (i) the pump also
resonates inside the cavity (doubly resonant case), which requires in
addition dispersion compensation at the pump spectral region, or (ii) the
cavity is transparent for the pump (singly resonant case), a case which is
free from the previous restriction and thus more amenable for
experimentation, at the expense of a higher threshold, as we will see. We
detail more the latter case in the appendix at the end of this paper. We
will limit here our analysis to non-chirped pumps as chirping requires a
more general treatment which will be presented elsewhere.

As the finesse of the cavity is assumed to be high, the intracavity signal
field operator $\hat{E}_{\mathrm{s}}$ can be written as a superposition of
cavity modes. Inside the $\chi ^{\left( 2\right) }$ crystal, which extends
from $z=-l/2$ to $z=+l/2$, one can write
\begin{equation}
\hat{E}_{\mathrm{s}}(z,t)=\sum\limits_{m}i\mathcal{E}_{\mathrm{s},m}\hat{s}%
_{m}(t)u_{m}\left( z\right) e^{-i\omega _{\mathrm{s},m}t}+\mathrm{H.c.}
\label{defEs}
\end{equation}%
where $\omega _{\mathrm{s},m}=\omega _{0}+m\Omega $, $\hat{s}_{m}\left(
t\right) $ is the annihilation operator for the $m$-th signal mode in the
interaction picture, verifying standard boson commutation relations%
\begin{equation}
\left[ \hat{s}_{m}\left( t\right) ,\hat{s}_{n}^{\dag }\left( t\right) \right]
=\delta _{m,n},  \label{boson_s}
\end{equation}%
$u_{m}\left( z\right) $ is the spatial profile of mode $m$, equal to $e^{ik_{%
\mathrm{s},m}z}$ in the case of ring cavities, while for linear cavities it
is equal to $\sin \left[ k_{\mathrm{s},m}\left( z+L/2\right) \right] $,
where
\begin{equation}
k_{\mathrm{s},m}=k\left( \omega _{\mathrm{s},m}\right) =\frac{n\left( \omega
_{\mathrm{s},m}\right) \omega _{\mathrm{s},m}}{c},  \label{defkm}
\end{equation}%
is the corresponding wavenumber. Finally $\mathcal{E}_{\mathrm{s},m}$ is the
single photon field amplitude, whose value depends on the type of cavity.
For a ring cavity
\begin{equation}
\mathcal{E}_{\mathrm{s},m}^{\left( \mathrm{ring}\right) }=\sqrt{\frac{\hbar
\omega _{\mathrm{s},m}}{2\varepsilon _{0}n\left( \omega _{\mathrm{s}%
,m}\right) A_{\mathrm{s}}L}},
\end{equation}%
where $A_{\mathrm{s}}$ is the transverse area of the signal field, while for
a linear cavity
\begin{equation}
\mathcal{E}_{\mathrm{s},m}^{\left( \mathrm{linear}\right) }=\sqrt{2}\mathcal{%
E}_{\mathrm{s},m}^{\left( \mathrm{ring}\right) }.
\end{equation}%
Note that we are writing the field as a superposition of plane waves, but
the treatment is approximately valid for Gaussian beams provided that the
crystal is placed at the beam waist and the Rayleigh length is much longer
than the crystal length $l$. In this case $A_{\mathrm{s}}=\pi w_{\mathrm{s}%
}^{2}$ with $w_{\mathrm{s}}$ the beam radius. Similarly we have $A_{\mathrm{p%
}}=\pi w_{\mathrm{p}}^{2}$ for the pump transverse mode.

The interaction Hamiltonian $\hat{H}_{\mathrm{I}}$ describing the parametric
interaction in the nonlinear crystal is given as usual by:
\begin{equation}
\hat{H}_{\mathrm{I}}=-A_{\mathrm{I}}{\textstyle\int\limits_{-l/2}^{+l/2}}%
\mathrm{d}z\left[ \hat{E}_{\mathrm{p}}\left( z,t\right) \hat{P}_{\mathrm{p}%
}\left( z,t\right) +\hat{E}_{\mathrm{s}}\left( z,t\right) \hat{P}_{\mathrm{s}%
}\left( z,t\right) \right] ,  \label{HIdef}
\end{equation}%
where $\hat{P}_{\mathrm{s}}\left( z,t\right) $ and $\hat{P}_{\mathrm{p}%
}\left( z,t\right) $ are the nonlinear electric polarization at signal and
pump frequencies, and $A_{\mathrm{I}}$ accounts for the effective area of
interaction corresponding to the three-mode overlapping integral across the
transverse plane and, for Gaussian beams, it is given by $A_{\mathrm{I}%
}^{-1}=A_{\mathrm{p}}^{-1}+2A_{\mathrm{s}}^{-1}$. The calculation of the
Hamiltonian depends on the type of configuration (singly- or doubly
resonant). Here we consider the simpler case of a doubly resonant SPOPO and
leave the details of the singly resonant case for the Appendix.

\subsubsection{Doubly resonant SPOPO}

In this case an expression for the intracavity pump field operator $\hat{E}_{%
\mathrm{p}}$ analogous to (\ref{defEs}), now centered around $2\omega _{0}$,
can be used and the following expression for $\hat{H}_{\mathrm{I}}$ in the
rotating wave approximation is obtained:
\begin{equation}
\begin{split}
\hat{H}_{\mathrm{I}}=2i\varepsilon _{0}\chi lA_{\mathrm{I}}\sum_{m,q}&
\mathcal{E}_{\mathrm{s},m}\mathcal{E}_{\mathrm{s},q}\mathcal{E}_{\mathrm{p}%
,m+q}f_{m,q}\times  \\
& \times \hat{s}_{m}^{\dag }\left( t\right) \hat{s}_{q}^{\dag }\left(
t\right) \hat{p}_{m+q}\left( t\right) +\mathrm{H.c.,}
\end{split}%
\end{equation}%
where $\chi $ is the relevant nonlinear susceptibility, $\hat{p}_{m}\left(
t\right) $ and $\hat{p}_{m}^{\dag }\left( t\right) $ are pump boson
operators ($m=0$ denotes the phase matched mode) verifying $\left[ \hat{p}%
_{m}\left( t\right) ,\hat{p}_{m}^{\dag }\left( t\right) \right] =\delta
_{m,n}$, and $\mathcal{E}_{\mathrm{p},m}$ is as $\mathcal{E}_{\mathrm{s},m}$
with the substitutions $\omega _{\mathrm{s},m}\rightarrow \omega _{\mathrm{p}%
,m}=2\omega _{0}+m\Omega $ and $A_{\mathrm{s}}\rightarrow A_{\mathrm{p}}$.
The phase-mismatching factor of the crystal, $f_{m,q}$, is given by:
\begin{equation}
f_{m,q}=\frac{\sin \phi _{m,q}}{\phi _{m,q}},  \label{fmq}
\end{equation}%
$\phi _{m,q}$ being the phase-mismatch angle:
\begin{equation}
\phi _{m,q}=\frac{1}{2}\left( k_{\mathrm{p},m+q}-k_{\mathrm{s},m}-k_{\mathrm{%
s},q}\right) l.  \label{phimq}
\end{equation}

Making use of the standard input-output formalism of optical cavities the
following set of Heisenberg equations for the signal annihilation operators $%
\hat{s}_{m}$ is derived straightforwardly:
\begin{equation}
\frac{d\hat{s}_{m}}{dt}=-\gamma _{\mathrm{s}}\hat{s}_{m}+\sqrt{2\gamma _{%
\mathrm{s}}}\hat{s}_{\mathrm{in},m}+\kappa \sum\nolimits_{q}f_{m,q}\hat{s}%
_{q}^{\dag }\hat{p}_{m+q},  \label{dsdr}
\end{equation}%
where the cavity damping rate $\gamma _{\mathrm{s}}$, or cavity linewidth,
is equal to $\frac{\Omega T_{\mathrm{s}}}{4\pi }$, $T_{\mathrm{s}}\ll 1$ is
the transmission factor of the single cavity mirror at which losses are
assumed to be concentrated, the coupling constant $\kappa $ is given by
\begin{equation}
\kappa =\mathcal{G}\chi l\frac{A_{\mathrm{I}}}{A_{\mathrm{s}}\sqrt{A_{%
\mathrm{p}}}}\left( \frac{\omega _{0}}{n_{0}L}\right) ^{3/2}\sqrt{\frac{%
\hbar }{\varepsilon _{0}}},  \label{ka}
\end{equation}%
and $\mathcal{G}$ is a factor depending on the geometry of the cavity that
amounts to $2$ for the ring cavity and to $\sqrt{2}$ for the linear cavity.

Analogously, the evolution equations for the pump annihilation operators $%
\hat{p}_{q}$ are:
\begin{equation}
\frac{d\hat{p}_{q}}{dt}=-\gamma _{\mathrm{p}}\hat{p}_{q}+\sqrt{2\gamma _{%
\mathrm{p}}}\hat{p}_{\mathrm{in},q}-\frac{\kappa }{2}\sum\nolimits_{m}f_{m,q}%
\hat{s}_{m}\hat{s}_{q-m},  \label{pump}
\end{equation}%
where $\gamma _{\mathrm{p}}$ is the cavity damping coefficient evaluated at
pump frequencies.

To get these simple equations, we have assumed that $\mathcal{E}_{\mathrm{s}%
,m}=\mathcal{E}_{\mathrm{s},0}$ $\forall m$, and neglected the dispersion of
the nonlinear susceptibility, which is a very good approximation as far as
the pulses bandwidth is not too large \cite{Seres}.
%pulses are in the 100 fs range.
The \textquotedblleft \textrm{in}" operators correspond to quantum fields
entering the cavity through the coupling mirror. We consider the case where
the input signal field is the vacuum, and the input pump field a coherent
state. We have therefore $\left\langle \hat{s}_{\mathrm{in},m}\left(
t\right) \right\rangle =0$, $\left\langle \hat{p}_{\mathrm{in},q}\left(
t\right) \right\rangle =p_{\mathrm{ext},q}$, and the following correlations
\begin{eqnarray}
\left\langle \hat{p}_{\mathrm{in},m}\left( t\right) ,\hat{p}_{\mathrm{in}%
,m^{\prime }}^{\dag }\left( t^{\prime }\right) \right\rangle
&=&\left\langle \hat{s}_{\mathrm{in},m}\left( t\right) ,\hat{s}_{\mathrm{in}%
,m^{\prime }}^{\dag }\left( t^{\prime }\right) \right\rangle   \notag \\
&=&\delta _{m,m^{\prime }}\delta \left( t-t^{\prime }\right) ,  \label{c1}
\end{eqnarray}%
with the notation $\left\langle \hat{a},\hat{b}\right\rangle =\left\langle
\left( \hat{a}-\left\langle \hat{a}\right\rangle \right) \left( \hat{b}%
-\left\langle \hat{b}\right\rangle \right) \right\rangle $, the rest of
correlations being null. The mean input field $p_{\mathrm{ext},q}$ is
related to the $\alpha _{q}$ and $P$ coefficients introduced in \eqref{Eext}
by:
\begin{equation}
p_{\mathrm{ext},q}=\sqrt{\frac{n_{0}A_{\mathrm{p}}P}{2\hbar \omega _{0}}}%
\alpha _{q}.  \label{pext}
\end{equation}

\subsubsection{Singly resonant SPOPO}

As detailed in the Appendix, in the case where only the signal field is
resonating into the cavity the evolution of the signal field annihilation
operators is given by an expression identical to (\ref{dsdr}), but now the
pump annihilation operators are given by
\begin{equation}
\begin{split}
\hat{p}_{m}(t)=& \mathcal{G}\sqrt{\frac{L}{c}}\left( \hat{p}_{\mathrm{in}%
,m}^{(+)}(t)+\hat{p}_{\mathrm{in},m}^{(-)}(t)\right) + \\
& -\mathcal{G}^{2}\frac{L}{c}\kappa \sum_{n}f_{n,m-n}\hat{s}_{n}(t)\hat{s}%
_{m-n}(t).
\end{split}
\label{srpump}
\end{equation}

We see that these operators contain two contributions. The first one
corresponds to the free-field part that is described by means of two
independent boson operators $\hat{p}_{\mathrm{in},m}^{(\pm )}(t)$ associated
to the fields impinging the cavity from both the directions labeled with the
superscripts $(\pm )$. We consider the unidirectional pumping case where the
input pump field propagating from left to right (labeled with $(+)$) is a
coherent state with a mean value of $\langle \hat{p}_{\mathrm{in}%
,q}^{(+)}(t)\rangle =p_{\mathrm{ext},q}$, while the input pump field
propagating form right to left (and labeled with $(-)$) is the vacuum $%
\langle \hat{p}_{\mathrm{in},q}^{(-)}(t)\rangle =0$. The mean input field, $%
p_{\mathrm{ext},q}$, is still given by Eq. (\ref{pext}), and the only
non-null correlations are:
\begin{equation}
\left\langle \hat{p}_{\mathrm{in},m}^{(\pm )}\left( t\right) ,\left[ \hat{p}%
_{\mathrm{in},n}^{(\pm )}\left( t^{\prime }\right) \right] ^{\dag
}\right\rangle =\delta _{m,n}\,\delta \left( t-t^{\prime }\right) .
\end{equation}

\subsection{The SPOPO below threshold}

Below threshold signal modes have a zero mean value, whereas the pump field
is characterized by a huge amplitude. One can therefore use a linearization
procedure for the quantum fluctuations, which amounts to setting $\hat{p}%
_{m+q}\rightarrow \left\langle \hat{p}_{m+q}\right\rangle $ in (\ref{dsdr}).
In the doubly resonant case $\left\langle \hat{p}_{m+q}\right\rangle =\sqrt{%
2/\gamma _{\mathrm{p}}}\left\langle \hat{p}_{\mathrm{in},m+q}\right\rangle =%
\sqrt{2/\gamma _{\mathrm{p}}}\,p_{\mathrm{ext},m+q}$, as given by (\ref{pump}%
), while in the singly resonant case $\left\langle \hat{p}%
_{m+q}\right\rangle =\mathcal{G}\sqrt{L/c}\left\langle \hat{p}_{\mathrm{in}%
,m+q}^{\left( +\right) }\right\rangle =\mathcal{G}\sqrt{L/c}\,p_{\mathrm{ext}%
,m+q}$, as given by (\ref{srpump}). In both cases the final equations for
the signal field anihilation operators are identical:
\begin{equation}
\frac{d\hat{s}_{m}}{dt}=-\gamma _{\mathrm{s}}\hat{s}_{m}+\sqrt{2\gamma _{%
\mathrm{s}}}\hat{s}_{\mathrm{in},m}+\gamma _{\mathrm{s}}\sigma
\sum\nolimits_{q}\mathcal{L}_{m,q}\hat{s}_{q}^{\dag },  \label{dsdtdouble}
\end{equation}%
where
\begin{equation}
\sigma =\sqrt{P/P_{0}},
\end{equation}%
is a pump amplitude parameter, and $P_{0}$ is an important scaling parameter
for the pump, which can be shown to be the threshold value for the pump in
the c.w. regime (single mode pump configuration). In the optimized
configuration for the pump focussing ($A_{\mathrm{p}}=A_{\mathrm{s}}/2$), it
is equal to:
\begin{equation}
P_{0}=\Pi _{0}\frac{\varepsilon _{0}c^{3}n_{0}^{2}T_{\mathrm{s}}^{2}}{%
2\left( \chi l\omega _{0}\right) ^{2}},  \label{P0}
\end{equation}%
\begin{table}[h]
\centering
\begin{tabular}{c|c|c}
\hline\hline
$\Pi_{0}$ & Doubly resonant & Singly resonant \\ \hline
Linear cavity & $T_{\mathrm{p}}/16$ & $1$ \\ \hline
Ring cavity & $T_{\mathrm{p}}/4$ & $1$ \\ \hline
\end{tabular}%
\caption{Coefficient for retrieving the cw threshold $P_{0}$ in the
different experimental situations considered.}
\label{SPOPOthreshold}
\end{table}
where $\Pi _{0}$ is given in Table \ref{SPOPOthreshold} for the singly or
doubly resonant configurations and ring or linear geometries. The
differences arise from the fact that the doubly resonant case presents an
intracavity pump power enhancement factor of $4/T_{\mathrm{p}}$ with respect
to the singly resonant case and the crystal is used twice in a linear cavity
as compared to the ring one\textbf{.} Hence the ratio $P_{0}^{\left( \mathrm{%
doubly}\right) }/P_{0}^{\left( \text{\textrm{singly}}\right) }$ equals the
(very small) pump transmission factor of the doubly resonant cavity. By way
of example, if we consider a singly resonant SPOPO \textbf{(}the geometry
does not matter\textbf{)} based on a BIBO crystal \cite{BIBO} with a
thickness of $l=100\,\mu$m and pumped at $0.4\,\mu$m, we
obtain reference irradiances $P_{0}$ of approximate values $14$, $344$, and $%
1400\,\mathrm{MW}\,\mathrm{cm}^{-2}$ for $T_{\mathrm{s}}=0.01$, $0.05$, and $0.1$,
respectively. For a typical pump beam radius of $70\,\mu$m these
irradiances lead to pump powers equal to $2$, $53$, and $212\,\mathrm{kW}$,
respectively.

The key point for the following analysis is the fact that, in equations (\ref%
{dsdtdouble}), the parametric coupling between the different signal modes is
\textit{linear}. It is characterized by a matrix $\mathcal{L}$, with matrix
elements:
\begin{equation}
\mathcal{L}_{m,q}=f_{m,q}\alpha _{m+q}=\frac{\sin \phi _{m,q}}{\phi _{m,q}}%
\alpha _{m+q}.  \label{Lmq}
\end{equation}%
When necessary, the phase mismatch angle $\phi _{m,q}$, Eq. \eqref{phimq}
can be computed using a Taylor expansion around $2\omega _{0}$ for the pump
wave vectors $k_{\mathrm{p},m}$ and around $\omega _{0}$ for the signal wave
vectors $k_{\mathrm{s},m}$,
\begin{equation}
\phi _{m,q}\simeq \beta _{1}\left( m+q\right) +\beta _{2\mathrm{p}}\left(
m+q\right) ^{2}-\beta _{2\mathrm{s}}\left( m^{2}+q^{2}\right) ,  \label{phi}
\end{equation}%
where
\begin{align}
\beta _{1}& =\frac{1}{2}\Omega \left( k_{\mathrm{p}}^{\prime }-k_{\mathrm{s}%
}^{\prime }\right) l,  \label{beta1} \\
\beta _{2\mathrm{p}}& =\frac{1}{4}\Omega ^{2}k_{\mathrm{p}}^{\prime \prime
}l,  \label{beta2p} \\
\beta _{2\mathrm{s}}& =\frac{1}{4}\Omega ^{2}k_{\mathrm{s}}^{\prime \prime
}l,  \label{beta2s}
\end{align}%
are dispersion coefficients, and $k^{\prime }$ and $k^{\prime \prime }$ are
the first and second derivatives of the wave vector with respect to
frequency. Note that the matrix $\mathcal{L}$ depends on the cavity
characteristics only through the free spectral range $\Omega $.

\section{SPOPO dynamics for the mean fields. Determination of the SPOPO
threshold}

Before calculating the quantum fluctuations of the SPOPO we analyze first
the dynamics of the mean values of the operators, which is obtained by
removing in Eq. \eqref{dsdr} the input noise terms and replacing the
operators by complex numbers:
\begin{equation}
\frac{ds_{m}}{dt}=-\gamma_{\mathrm{s}}s_{m}+\gamma_{\mathrm{s}}\sigma
\sum\limits_{q}\mathcal{L}_{m,q}s_{q}^{\ast}.  \label{dsmdtclas}
\end{equation}

The solution to Eqs. \eqref{dsmdtclas} is of the form
\begin{equation}
s_{m}\left( t\right) =S_{k,m}e^{\lambda_{k}t},
\end{equation}
where $k$ is an index labelling the different solutions, and the parameters $%
S_{k,m}$ and $\lambda_{k}$ obey the following eigenvalue equation:
\begin{equation}
\lambda_{k}S_{k,m}=-\gamma_{\mathrm{s}}S_{k,m}+\gamma_{\mathrm{s}}\sigma
\sum\limits_{q}\mathcal{L}_{m,q}S_{k,q}^{\ast}.  \label{LkSkm}
\end{equation}

As matrix $\mathcal{L}$ is both self-adjoint and real, its eigenvalues $%
\Lambda_{k}$ and eigenvectors $\vec{L}_{k}$, of components $L_{k,m}$ defined
by
\begin{equation}
\Lambda_{k}L_{k,m}=\sum\limits_{q}\mathcal{L}_{m,q}L_{k,q}  \label{eigeqL}
\end{equation}
are all real. As $\gamma_{\mathrm{s}}$ and $\sigma$ are also real, it is
evident that two sets of solutions to Eqs. \eqref{LkSkm} exist, namely $%
S_{k,m}^{\left( +\right) }=L_{k,m}$ and $S_{k,m}^{\left( -\right)}=iL_{k,m}$%
, with corresponding eigenvalues:
\begin{equation}
\lambda_{k}^{\left( \pm\right) }=\gamma_{\mathrm{s}}\left(-1\pm\sigma%
\Lambda_{k}\right) .  \label{lambda}
\end{equation}
Let us label by index $k=0$ the solution of maximum value of $|\Lambda_{k}|$%
. When $\sigma\left\vert \Lambda_{0}\right\vert <1$, all the rates $%
\lambda_{k}^{\pm}$ are negative, which implies that the null solution for
the steady state signal field is stable. For simplicity of notation, we will
take $\Lambda_{0}$ positive in the following, which is a common situation as
shown below \cite{lam0}. Hence $\lambda_{0}^{\left( +\right) }$ is the
largest eigenvalue and $\lambda _{0}^{\left( +\right) }=0$ sets the SPOPO
oscillation threshold, which then occurs when the pump parameter $\sigma$
takes the value $1/\Lambda_{0}$, i.e. for a pump irradiance $P=P_{\mathrm{thr%
}}$ equal to:
\begin{equation}
P_{\mathrm{thr}}=P_{0}/\Lambda_{0}^{2}.  \label{Pthr}
\end{equation}
The exact value of $\Lambda_{0}$, and therefore of the SPOPO threshold,
depends on the exact shape of the phase matching curve and on the exact
spectrum of the pump laser. As will be shown in Sec. V, the theoretical
SPOPO threshold can be extremely low, of the order of the cw single mode
threshold divided by the number of pump modes.

Let us now define the normalized amplitude pumping rate $r$ by
\begin{equation}
r=\sqrt{P/P_{\mathrm{thr}}},
\end{equation}
or $r=\sigma\Lambda_{0}$, so that the threshold occurs at $r=1$. The
eigenvalues $\lambda_{k}$ become%
\begin{equation}
\lambda_{k}^{\left( \pm\right) }=\gamma_{\mathrm{s}}\left( -1\pm r\frac{%
\Lambda_{k}}{\Lambda_{0}}\right) .  \label{lambdak_r}
\end{equation}

We will call supermodes the set of $S_{k,m}$ values for a given $k$, which
corresponds physically to the different spectral components of the signal
field, and critical supermode $S_{k=0,m}^{\left( +\right) }$, the one
associated with $\lambda_{0}^{\left( +\right) }$, which is the eigenvalue
changing its sign at threshold. Above threshold, this critical mode will be
the ``lasing" one, i.e. the one having a non-zero mean amplitude when $r>1$.
Note that the supermodes are independent of pump (Eq. \ref{eigeqL}),but not
the eigenvalues (Eq. \ref{lambda}).

We note that the supermode in quadrature with respect to the critical one, $%
S_{0}^{\left( -\right) }=iS_{0}^{\left( +\right) }$, has an associated
eigenvalue $\lambda_{0}^{\left( -\right) }=-2\gamma_{\mathrm{s}}$ at
threshold, Eq. (\ref{lambdak_r}) with $r=1$, which is the lowest eigenvalue
below- or at threshold. This property is obvious: Should $%
\lambda_{k}^{\left( -\right) }<-2\gamma_{\mathrm{s}}$ for some $k$, then $r%
\frac{\Lambda_{k}}{\Lambda_{0}}$ should be larger than $1$, what is
incompatible with the fact that $\left\vert \frac{\Lambda_{k}}{\Lambda_{0}}%
\right\vert <1$ by definition and the condition $r<1$. The fact that there
exists an eigenvector whose damping rate ($\lambda_{0}^{\left( -\right)
}=-2\gamma_{\mathrm{s}}$ at threshold) is twice that of the passive cavity
has important consequences on the squeezing properties of the SPOPO, as it
occurs in other OPO configurations \cite{DOPOsoliton}.

\section{Quantum fluctuations of the SPOPO below threshold}

\subsection{Fluctuation spectrum for the supermodes}

We can now determine the quantum fluctuations of the signal field in a SPOPO
below threshold. Let us introduce the following operators:
\begin{align}
\hat{S}_{k}(t) & =\sum\limits_{m}L_{k,m}\hat{s}_{m}(t),  \label{Sk} \\
\hat{S}_{\mathrm{in},k}(t) & =\sum\limits_{m}L_{k,m}\hat{s}_{\mathrm{in}%
,m}(t).  \label{Sink}
\end{align}
As $\vec{L}_{k}\cdot\vec{L}_{k^{\prime}}\equiv\sum\nolimits_{m}L_{k,m}L_{k^{%
\prime},m}=\delta_{k,k^{\prime}}$, one has trivially
\begin{align*}
\left[ \hat{S}_{k}(t),\hat{S}_{k^{\prime}}^{\dagger}(t)\right] &
=\delta_{k,k^{\prime}}, \\
\left[ \hat{S}_{\mathrm{in},k}(t),\hat{S}_{\mathrm{in},k^{\prime}}^{\dagger
}(t^{\prime})\right] & =\delta_{k,k^{\prime}}\delta(t-t^{\prime}),
\end{align*}
and the correlation%
\begin{equation}
\left\langle \hat{S}_{\mathrm{in},k}(t),\hat{S}_{\mathrm{in}%
,k^{\prime}}^{\dagger}(t^{\prime})\right\rangle
=\delta_{k,k^{\prime}}\delta\left( t-t^{\prime}\right) ,
\end{equation}
as well. Hence $\hat{S}_{k}$ and $\hat{S}_{\mathrm{in},k}$ are the
annihilation operators of a combination of signal modes of different
frequencies, which are the eigenmodes of the linearized evolution equation (%
\ref{dsmdtclas}). The corresponding creation operator applied to the vacuum
state creates a photon in a single supermode which globally describes the
frequency comb, or train of pulses. Analogously one can define supermode
output operators%
\begin{equation}
\hat{S}_{\mathrm{out},k}(t)=\sum\limits_{m}L_{k,m}\hat{s}_{\mathrm{out}%
,m}(t),  \label{Soutk}
\end{equation}
where the output boson operator $\hat{s}_{\mathrm{out},m}\left( t\right) $
relates to the intracavity and input boson operators through the usual
input-output relation of high finesse optical cavities,%
\begin{equation}
\hat{s}_{\mathrm{out},m}\left( t\right) = -\hat{s}_{\mathrm{in},m}\left(
t\right) + \sqrt{2\gamma_{\mathrm{s}}}\hat{s}_{m}\left( t\right) .
\label{inout-t}
\end{equation}

One can then write:
\begin{equation}
\frac{d}{dt}\hat{S}_{k}=-\gamma_{\mathrm{s}}\hat{S}_{k}+\gamma_{\mathrm{s}%
}\sigma\Lambda_{k}\hat{S}_{k}^{\dagger}+\sqrt{2\gamma_{\mathrm{s}}}\hat {S}_{%
\mathrm{in},k}.  \label{d3}
\end{equation}
Let us now define quadrature hermitian operators $\hat{S}_{k}^{\left(
\pm\right) }$ by:
\begin{align}  \label{quad}
\hat{S}_{k}^{\left( +\right) } & =\hat{S}_{k}+\hat{S}_{k}^{\dagger}, \\
\hat{S}_{k}^{\left( -\right) } & =-i\left( \hat{S}_{k}-\hat{S}%
_{k}^{\dagger}\right) ,
\end{align}
and analogously for $\hat{S}_{\mathrm{in},k}$ and $\hat{S}_{\mathrm{out},k}$%
, which obey the following equations:
\begin{equation}
\frac{d}{dt}\hat{S}_{k}^{\left( \pm\right) }=\lambda_{k}^{\left( \pm\right) }%
\hat{S}_{k}^{\left( \pm\right) }+\sqrt{2\gamma_{\mathrm{s}}}\hat{S}_{\mathrm{%
in},k}^{\left( \pm\right) },  \label{d4}
\end{equation}
with $\lambda_{k}^{\left( \pm\right) }$ given by Eq. (\ref{lambda}). These
relations enable us to determine the intracavity quadrature operators in the
Fourier domain $\tilde{S}_{k}^{\left( \pm\right) }(\omega)$
\begin{equation}
i\omega\tilde{S}_{k}^{\left( \pm\right) }(\omega)=\lambda_{k}^{\left(
\pm\right) }\tilde{S}^{\pm}(\omega)+\sqrt{2\gamma_{\mathrm{s}}}\tilde {S}_{%
\mathrm{in},k}^{\left( \pm\right) }(\omega).  \label{d2}
\end{equation}

Finally, the usual input-output relation on the coupling mirror (\ref%
{inout-t}), which can be written as%
\begin{equation}
\tilde{s}_{\mathrm{out},m}(\omega)=-\tilde{s}_{\mathrm{in},m}(\omega )+\sqrt{%
2\gamma_{\mathrm{s}}}\tilde{s}_{m}(\omega),  \label{inout}
\end{equation}
being $\tilde{s}_{\mathrm{out},m}(\omega)$ the Fourier transform of the
output boson operator $\hat{s}_{\mathrm{out},m}\left( t\right) $, extends by
linearity to any supermode operator as the mirror is assumed to have a
transmission independent of the mode frequency. One then obtains the
following expression for the quadrature component in Fourier space of any
signal supermode,
\begin{align}
\tilde{S}_{\mathrm{out},k}^{\left( \pm\right) }(\omega) & =v_{k}^{\left(
\pm\right) }\left( \omega\right) \tilde{S}_{\mathrm{in},k}^{\left(
\pm\right) }(\omega),  \label{d5} \\
v_{k}^{\left( \pm\right) }\left( \omega\right) & =\frac{\gamma _{\mathrm{s}%
}\left( 1\pm r\Lambda_{k}/\Lambda_{0}\right) -i\omega}{\gamma_{\mathrm{s}%
}\left( -1\pm r\Lambda_{k}/\Lambda_{0}\right) +i\omega}.  \label{vk}
\end{align}
One has also, for the operators in Fourier space:
\begin{eqnarray}  \label{SinSin}
\left\langle \tilde{S}_{\mathrm{in},k}^{\left(a\right) }\left(
\omega_1\right) \tilde{S}_{\mathrm{in},l}^{\left( b\right) }\left(
\omega_{2}\right) \right\rangle &=&\frac{\eta^{(a,b)}}{2\pi}%
\delta_{kl}\delta\left( \omega_1+\omega_{2}\right) \\
a&=&\pm,b=\pm
\end{eqnarray}
with $\eta^{(+,+)}=\eta^{(-,-)}=1$ and $\eta^{(+,-)}=-\eta^{(-,+)}=i$.

\subsection{Homodyne detection}

The variances of the quadrature operators can be measured using the usual
balanced homodyne detection scheme: the local oscillator (LO) is in the
present case a coherent mode-locked multimode field $E_{\mathrm{L}}\left(
t\right) $ having the same repetition rate as the pump laser:
\begin{align}
E_{\mathrm{L}}\left( t\right) & =E_{\mathrm{L}}^{-}\left( t\right) +E_{%
\mathrm{L}}^{+}\left( t\right) \\
E_{\mathrm{L}}^{-}\left( t\right) & =i\epsilon_{\mathrm{L}%
}\sum_{m}e_{m}e^{-i\omega_{\mathrm{s},m}t}, \\
E_{\mathrm{L}}^{+}\left( t\right) & =\left[ E_{\mathrm{L}}^{-}\left(
t\right) \right] ^{\ast},
\end{align}
where $\sum_{m}|e_{m}|^{2}=1$, and $\epsilon_{\mathrm{L}}$ is the LO field
total amplitude factor. The output signal exiting the SPOPO, $\hat {E}_{s,%
\mathrm{out}}\left( t\right) $, is combined with $E_{\mathrm{L}}\left(
t\right) $ in a 50\%--50\% beam splitter, the intensity of the two output
ports is measured using photodiodes of unity quantum efficiency, and their
difference constitutes the homodyne signal. Writing
\begin{align*}
\hat{E}_{s,\mathrm{out}}\left( t\right) & =\hat{E}_{s,\mathrm{out}%
}^{-}\left( t\right) +\hat{E}_{s,\mathrm{out}}^{+}\left( t\right) , \\
\hat{E}_{s,\mathrm{out}}^{-}\left( t\right) & =i\mathcal{E}_{\mathrm{out}%
}\sum\nolimits_{n}\hat{s}_{\mathrm{out},m}\left( t\right) e^{-i\omega _{%
\mathrm{s},m}t}, \\
\hat{E}_{s,\mathrm{out}}^{+}\left( t\right) & =\left[ \hat{E}_{s,\mathrm{out}%
}^{-}\left( t\right) \right] ^{\dag},
\end{align*}
where $\mathcal{E}_{\mathrm{out}}$ is a proportionality constant. If
sufficiently fast detectors were used the measurement would give an
instantaneous signal represented by the operator%
\begin{equation*}
\hat{\imath}\left( t\right) =\frac{1}{\epsilon_{\mathrm{L}}\mathcal{E}_{%
\mathrm{out}}}\left[ E_{\mathrm{L}}^{-}\left( t\right) \hat {E}_{s,\mathrm{%
out}}^{+}\left( t\right) +E_{\mathrm{L}}^{+}\left( t\right) \hat{E}_{s,%
\mathrm{out}}^{-}\left( t\right) \right] .
\end{equation*}
When detectors are not so fast (we are considering inter-pulse separations
on the order of few $\mathrm{ns}$) they average over many pulses along their
response time $\tau_{\mathrm{d}}$ and $\hat{\imath}$ must be substituted by $%
\hat{\imath}_{\mathrm{H}}\left( t\right) =\frac{1}{\tau_{\mathrm{d}}}%
\int_{t-\tau_{\mathrm{d}}/2}^{t+\tau_{\mathrm{d}}/2}dt^{\prime}\hat{\imath}%
\left( t^{\prime}\right) $, which can be very well approximated by%
\begin{equation}
\hat{\imath}_{\mathrm{H}}\left( t\right) =\sum\limits_{m}\left[ e_{m}\hat{s}%
_{\mathrm{out},m}^{\dag}\left( t\right) +e_{m}^{\ast}\hat {s}_{\mathrm{out}%
,m}\left( t\right) \right] ,  \label{iH}
\end{equation}
where we considered that $\tau_{\mathrm{d}}\gg2\pi/\Omega$ and used that $%
\hat{s}_{\mathrm{out},m}\left( t\right) $ and $\hat{s}_{\mathrm{out}%
,m}^{\dag}\left( t\right) $ vary little during the time $\tau_{\mathrm{d}}$
\cite{input-output}, what roughly requires that $\tau_{\mathrm{d}}\ll
\gamma_{\mathrm{s}}^{-1}$. Note that this case, namely $2\pi/\Omega\ll \tau_{%
\mathrm{d}}\ll\gamma_{\mathrm{s}}^{-1}$, is sensible as $\gamma _{\mathrm{s}%
}^{-1}=2T_{\mathrm{s}}^{-1}\left( 2\pi/\Omega\right) $, where $T_{\mathrm{s}%
}\ll1$ is the transmission factor of the single cavity mirror at which
signal losses are assumed to be concentrated. Then operator $\hat {\imath}_{%
\mathrm{H}}$ (\ref{iH}) represents the outcome of a balanced homodyne
detection that uses as a local oscillator a modelocked laser with the same
repetition rate as the SPOPO and with spectral components given by $e_{m}$.
The variance of $\hat{\imath}_{\mathrm{H}}$ measures then the fluctuations
of the projection of the output field on the local oscillator.

\subsubsection{Perfect mode matching case}

When the coefficients $e_{m}$ of the LO field spectral decomposition are
equal, apart from a global phase $\phi_{\mathrm{L}}$, to the coefficients $%
L_{k,m}$ of the $k$-th supermode, $e_{m}=e^{i\phi_{\mathrm{L}}}L_{k,m}$, one
measures, according to (\ref{iH}), a photocurrent difference proportional to
\begin{align}
\hat{\imath}_{\mathrm{H}}\left( t\right) & =e^{i\phi_{\mathrm{L}}}\hat {S}_{%
\mathrm{out},k}^{\dag}(t)+e^{-i\phi_{\mathrm{L}}}\hat{S}_{\mathrm{out},k}(t)
\notag \\
& =\hat{S}_{\mathrm{out},k}^{\left( +\right) }(t)\cos\phi_{\mathrm{L}}+\hat{S%
}_{\mathrm{out},k}^{\left( -\right) }(t)\sin\phi_{\mathrm{L}}.
\end{align}

The two following variances, depending on the local oscillator phase value $%
\phi_{\mathrm{L}}$, are measured (see next subsection for the
demonstration),
\begin{align}
V_{k}^{\left( -\right) }\left( \omega\right) & =v_{k}^{\left( -\right)
}\left( \omega\right) v_{k}^{\left( -\right) }\left( -\omega\right) =\frac{%
\gamma_{\mathrm{s}}^{2}\left( 1-r\Lambda_{k}/\Lambda_{0}\right)
^{2}+\omega^{2}}{\gamma_{\mathrm{s}}^{2}\left( 1+r\Lambda_{k}/\Lambda
_{0}\right) ^{2}+\omega^{2}},  \label{VHgena} \\
V_{k}^{\left( +\right) }\left( \omega\right) & =v_{k}^{\left( +\right)
}\left( \omega\right) v_{k}^{\left( +\right) }\left( -\omega\right)
=\frac1{V_{k}^{\left( -\right) }\left( \omega\right)},  \label{VHgenb}
\end{align}
where $v_{k}^{\left(\pm\right) }$ are given in (\ref{vk}). Equations (\ref%
{VHgena},\ref{VHgenb}) show that the device produces, as expected, a minimum
uncertainty state and that quantum noise reduction below the standard
quantum limit (equal here to $1$) is achieved for any supermode
characterized by a non-zero $\Lambda_{k}$ value. Clearly which quadrature is
squeezed depends on the sign of $\Lambda_{k}/\Lambda_{0}$, so that when
positive, it is $\hat{S}_{k}^{\left( -\right) }$ the squeezed quadrature
(phase-quadrature squeezing) and vice-versa (amplitude-quadrature
squeezing). The smallest fluctuations are obtained close to threshold ($r=1$%
) and at zero Fourier frequency ($\omega=0$):
\begin{equation}
\left( V_{k}\right) _{\min}=\left( \frac{\Lambda_{0}-|\Lambda_{k}|}{%
\Lambda_{0}+|\Lambda_{k}|}\right) ^{2}  \label{Squeeze}
\end{equation}
In particular, if one uses as the local oscillator a copy of the critical
mode $k=0$ (identical to the one oscillating just above the threshold $r=1$)
one then gets perfect squeezing just below threshold and at zero noise
frequency, just like in the c.w. single mode case. But modes of $k\neq0$ may
be also significantly squeezed, provided that $|\Lambda_{k}/\Lambda_{0}|$ is
not much different from $1$. In the next Section we analyze the behavior of
the squeezing levels just described.

Our multi-mode approach of the problem has therefore allowed us to extract
from all the possible linear combinations of signal modes the ones in which
the quantum properties are concentrated.

\subsubsection{General case}

As always in quantum optics, the measurement of a high degree of squeezing
in SPOPOs requires the use of a mode matched LO, namely of spectral
components $e_{m}=e^{i\phi _{\mathrm{L}}}L_{k,m}$. It is not always an easy
task and was recognized in \cite{Grangier} as the main experimental
limitation in pulsed squeezing. With the present ultrashort pulses, one can
use pulse shaping techniques with the help of dispersive elements and
programmable phase modulators \cite{Jiang2007,Huang2008,Supradeepa2008}. In
view of future experiments, it is important to determine the noise levels
measured using a LO of arbitrary shape, in order to know the accuracy with
which the perfectly modematched LO must be approached using pulse shaping
techniques. We derive in this section the noise spectrum for a LO of
arbitrary shape, that we will use in section VI.

Let us define the projections of the LO frequency comb onto the supermodes $%
L_{k}$, as%
\begin{equation}
d_{k}=\sum\limits_{m}L_{k,m}e_{m}.  \label{betak}
\end{equation}
This expression can be inverted to yield%
\begin{equation}
e_{m}=\sum\limits_{k}L_{k,m} d_{k},  \label{emetak}
\end{equation}
where the well known result $\sum\nolimits_{k}L_{k,m}L_{k,n}=\delta_{m,n}$
involving the elements of a basis has been used. Substitution of Eq. (\ref%
{emetak}) into Eq. (\ref{iH}) yields%
\begin{equation}
\hat{\imath}_{\mathrm{H}}\left( t\right) =\sum\limits_{k}\left[ \mathrm{Re}%
\left( d_{k}\right) \hat{S}_{\mathrm{out},k}^{\left( +\right) }(t)+\mathrm{Im}%
\left( d_{k}\right) \hat{S}_{\mathrm{out},k}^{\left( -\right) }(t)\right] ,
\end{equation}
where the quadrature operators, Eq. (\ref{quad}), have been used.

The noise variance spectrum associated to $\hat{\imath}_{\mathrm{H}}\left(
t\right) $, $V\left( \omega\right) $, can be computed as%
\begin{equation}
V\left( \omega\right)=\int\nolimits_{-\infty}^{+\infty}d\tau \left\langle
\hat{\imath}_{\mathrm{H}}\left( t\right) \hat{\imath }_{\mathrm{H}}\left(
t+\tau\right) \right\rangle e^{-i\omega\tau}.
\end{equation}

Using Eqs. \eqref{d5}, \eqref{vk}, \eqref{SinSin} and \eqref{iH}, it is
finally equal to: %\begin{widetext}
\begin{align}  \label{Squeezingspectrum}
V\left(\omega\right)=\sum_{k} &\left\{ \left( \mathrm{\mathrm{Re}}%
d_{k}\right)^{2}
v_{k}^{\left(+\right)}\left(\omega\right)v_{k}^{\left(+\right)}\left(-\omega%
\right)+ \right.  \notag \\
&\left. + \left(\mathrm{\mathrm{Im}}d_{k}\right)^{2}
v_{k}^{\left(-\right)}\left(\omega\right)v_{k}^{\left(-\right)}\left(-\omega%
\right)+ \right.  \notag \\
&\left.+i\,\mathrm{\mathrm{Re}}d_{k}\,\mathrm{\mathrm{Im}}d_{k} \left[
v_{k}^{\left(+\right)}\left(\omega\right)v_{k}^{\left(-\right)}\left(-\omega%
\right)+ \right.\right.  \notag \\
&\left.\left. -
v_{k}^{\left(-\right)}\left(\omega\right)v_{k}^{\left(+\right)}\left(-\omega%
\right) \right] \right\}.
\end{align}
%\end{widetext}
Equation (\ref{Squeezingspectrum}) gives the general expression of the
squeezing spectrum corresponding to a generic LO defined by its supermodal
amplitudes $d_{k}$ given by Eq. \eqref{betak}. When the LO is proportional
to the supermode labeled by $k$, say $e_{m}=e^{i\phi_{\mathrm{L}}}L_{k,m}$,
and $\phi_{\mathrm{L}}=0,\pi/2$, the two special quadratures (\ref{quad})
are selected and the results (\ref{VHgena}) and (\ref{VHgenb}) are recovered.

\section{Diagonalization of the matrix $\mathcal{L}$: analytical and
numerical results}

We have seen that all the properties of the SPOPO are directly related to
the series of eigenvalues $\Lambda_k$, depending both on the phase matching
properties of the crystal and pump spectral characteristics. We will
consider now in more detail the characteristics of these eigenvalues.

\begin{figure}[t!]
\centering
\includegraphics[width=0.3\textwidth]{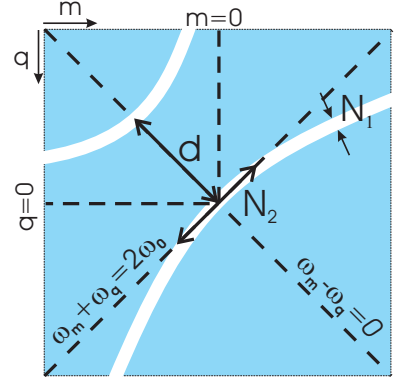}
\caption{(Color online) Sketch of the phase-matching matrix $f_{m,q}$ in the space of the
integer numbers $\{m,q\}$ corresponding to frequencies $\{\protect\omega_{m},%
\protect\omega_{q}\}$. In white are represented the regions of optimal
phase-matching while in blue the unmatched regions.}
\label{Legenda}
\end{figure}

Fig. \ref{Legenda} shows schematically the typical appearance of the
phase-mismatch matrix $f_{m,q}=\frac{\sin\phi_{m,q}}{\phi_{m,q}}$ (Eq. %
\eqref{phi}), for a typical configuration (see Section VI for details and
real examples). It has the shape of a hyperbola, whose branches, of width $%
N_{1}$, display a minimum distance between them called $d$. Another relevant
quantity is the ``width" $N_{2}$ marked in the figure. These quantities will
be useful for determining the Gaussian limit, that we will consider in the
next section.

The matrix $\mathcal{L}$ is the product of $f_{m,q}$ with the pump spectrum $%
\alpha_{m+q}$. As the last quantity is constant for $m+q=\mathrm{constant}$,
the pump selects a portion of matrix $f_{m,q}$, roughly given by the
intersection of $f_{m,q}$ with a straight band oriented along the direction $%
m+q=2m_{\max}$, where $m_{\max}$ corresponds to the maximum of the pump
spectrum.

\subsection{Analytical approach}

When the pump spectrum is not very broad, the resulting nonzero matrix
elements of $\mathcal{L}$ are confined within an ``ellipse" whose principal
axes are oriented along the directions $m+q=0$ and $m-q=0$. In this case one
can forget the secondary maxima of the sinc function and use a Gaussian
approximation for the phase-matching matrix:
\begin{equation}
f_{m,q}=e^{-\frac{1}{2}\left( \frac{m+q}{N_{1}}\right) ^{2}}e^{-\frac{1}{2}%
\left( \frac{m-q}{N_{2}}\right) ^{2}},  \label{fapprox}
\end{equation}
The form for $f_{m,q}$ follows from the approximations $\frac{\sin x}{x}%
\simeq e^{-\frac{x^{2}}{\eta_1}}$ and $\frac{\sin x^{2}}{x^{2}}\simeq e^{-%
\frac{x^{2}}{\eta_2}}$ \cite{Polish}, where the parameters $\eta_1$
and $\eta_2$ can be opportunely chosen so that the results from the
diagonalization of the coupling matrix obtained from \eqref{fapprox} match
optimally to the results of the numerical diagonalization of $\mathcal{L}%
_{m,q}$. By choosing them as $\eta_1=5$ and $\eta_2=12$, the following
expressions for the widths $N_{1}$ and $N_{2}$ can be obtained:
\begin{align}
N_{1}&=\frac{\sqrt{5/2}}{\left\vert \beta_{1}\right\vert },  \label{delta1}
\\
N_{2}&=\frac{2\sqrt{3}}{\sqrt{\left\vert \beta_{2\mathrm{s}}\right\vert }}.
\label{delta2}
\end{align}
where the $\beta$ coefficients are the ones introduced in \eqref{beta1}, %
\eqref{beta2p}, \eqref{beta2s}.

Let us assume in addition that the pump has a Gaussian spectrum centered at $%
2 \omega_0$:
\begin{equation}
\alpha_{m}=\pi^{-\frac {1}{4}}N_{\mathrm{p}}^{-\frac{1}{2}}e^{-\frac{1}{2}%
\left( \frac {m}{N_{\mathrm{p}}}\right) ^{2}}
\end{equation}
where $\sum\left\vert \alpha_{m}\right\vert^{2}=1$ and $N_{\mathrm{p}%
}=\left( \Omega\tau_{\mathrm{p}}\right) ^{-1}$ is a measure of the number of
pump modes. The matrix elements $\mathcal{L}_{m,q}$ are then equal to:
\begin{equation}
\mathcal{L}_{m,q}=e^{-\frac{1}{2}\left( \frac{m+q}{N_{1}}\right) ^{2}}e^{-%
\frac{1}{2}\left( \frac{m-q}{N_{2}}\right) ^{2}}\frac{1}{\pi^{1/4}\sqrt{N_{%
\mathrm{p}}}}e^{-\frac{1}{2}\left( \frac{m+q}{N_{\mathrm{p}}}\right) ^{2}}.
\label{Lapprox}
\end{equation}

The Gaussian approximation \eqref{Lapprox} is correct as far as the
interplay between the pump spectrum and the phase-matching is opportune.
More quantitatively speaking, we have to demand that the two branches of the
hyperbola in Fig. \ref{Legenda} are sufficiently separated each other, which
corresponds to require that $d\gg N_{1}$ (we will consider $d\gtrsim 10N_{1}$%
), and the pump width is sufficiently smaller than the width of
phase-matching function along the direction $m-q=0$, which corresponds to
the condition $N_{\mathrm{p}}\lesssim N_{1}$. These conditions lead to the
following bounds:
\begin{equation*}
\frac{\beta _{1}^{2}}{\left\vert \beta _{2\mathrm{p}}-\frac{1}{2}\beta _{2%
\mathrm{s}}\right\vert }>20,\text{ }2\left\vert \beta _{1}\right\vert N_{%
\mathrm{p}}<1.
\end{equation*}%
In terms of the crystal parameters and of the pump pulse duration these
validity limits can be cast as
\begin{align}
l& >20\frac{\left\vert k_{\mathrm{p}}^{\prime \prime }-\frac{1}{2}k_{\mathrm{%
s}}^{\prime \prime }\right\vert }{\left( k_{\mathrm{p}}^{\prime }-k_{\mathrm{%
s}}^{\prime }\right) ^{2}},  \label{taupmin} \\
\tau _{\mathrm{p}}& >\left\vert k_{\mathrm{p}}^{\prime }-k_{\mathrm{s}%
}^{\prime }\right\vert l.  \label{lmin}
\end{align}%
For a BIBO crystal under typical conditions the above inequalities read $%
l>0.2\,\mu$m (hence it is not a serious condition) and $\tau _{\mathrm{p}}>400\,\mathrm{fs%
}\times l/\mathrm{mm}$. Hence, for a crystal length $l=1\,\mathrm{mm}$, $%
\tau _{\mathrm{p}}$ should be larger than $400$ fs in order that the
Gaussian approximation is valid, while for $l=0.1$ mm the condition is
met just for $\tau _{\mathrm{p}}>40$ fs. We note that condition (\ref%
{taupmin}) means that the pump duration should be longer than the temporal
walk-off between signal and pump modes along their propagation inside the
crystal.

The eigenvalues of such a matrix turn out to have a simple analytical
expression at the continuous limit, i.e. when one can replace in (\ref%
{eigeqL}) the sum by an integral, so that:
\begin{equation}
\Lambda _{k}L_{k}\left( m\right) =\int dq\mathcal{L}\left( m,q\right)
L_{k}\left( q\right) ,  \label{diagLint}
\end{equation}%
(note that we changed the notation from indices to arguments).

The eigenvalues are given by
\begin{equation}  \label{LamGauss}
\Lambda_{k}=\Lambda_{0}\rho^{k},
\end{equation}
where
\begin{align}  \label{Lam0}
\Lambda_{0}&=\pi^{1/4}\sqrt{2N_{\mathrm{p}}}\sqrt{\frac{\tau_{\mathrm{p}}^{2}%
} {\tau_{1}^{2}+\tau_{\mathrm{p}}^{2}}}, \\
\label{rhoGauss}
\rho&=-1+2\sqrt{\frac{\tau_{2}^{2}}{\tau_{1}^{2}+\tau_{\mathrm{p}}^{2}}},
\end{align}
and
\begin{equation}
\tau_{1}=\frac{\left\vert k_{\mathrm{p}}^{\prime}-k_{\mathrm{s}}^{\prime
}\right\vert l}{\sqrt{10}},\;\tau_{2}=\frac{\sqrt{\left\vert k_{\mathrm{s}%
}^{\prime\prime}\right\vert l\ }}{4\sqrt{3}},  \label{tau12_Gauss}
\end{equation}
$\Lambda_0$ and $\rho$ are given in the limit $\tau_{2}\ll\tau_{1}$, which
holds unless the crystal length $l<0.1\mu$m; hence $\rho$ is very close to $-1$. Equation (\ref{LamGauss})
corresponds then to an alternating geometric progression of ratio $\rho$,
whose first element $\Lambda_{0}$ is positive.

The eigenvectors are similar to the well-known Hermite-Gauss $\mathrm{TEM}%
_{pq}$ transverse modes. They are given by%
\begin{equation}  \label{LkGauss}
L_{k,m}=\frac{1}{\sqrt{k!2^{k}\sqrt{\pi}N_{\mathrm{s}}}} e^{-\frac{1}{2}%
\left(\frac{m}{N_{\mathrm{s}}}\right)^{2}} H_{k}\left(\frac{m}{N_{\mathrm{s}}%
}\right),
\end{equation}
where $H_{k}$ is the Hermite polynomial of order $k$, and $N_{\mathrm{s}}$
is the number of signal modes. Hermite-Gauss functions being simply
proportional to their Fourier transforms, their temporal shape is exactly
the same as their spectral shape. The pulse duration of the zeroth mode, $%
\tau_{\mathrm{s}}$, is given by
\begin{equation}
\tau_{\mathrm{s}}^{2}=2\tau_{2}\sqrt{\tau_{1}^{2}+\tau_{\mathrm{p}}^{2}}.
\label{taus_Gauss}
\end{equation}
Under typical conditions \cite{ordersofmagnitude} the times $\tau_1$ and $%
\tau_2$ are on the order of $\tau_{1}\sim100$ fs and $\tau_{2}\sim5$ fs for a crystal
length $l=1$ mm, and in general $%
\tau_{2}\ll\tau_{1}$ whenever $l\gtrsim0.1\,\mu$m. Note that the
condition (\ref{taupmin}) implies that $\tau_{\mathrm{p}}^{2}\gg\tau_{1}^{2}$%
, so that $\Lambda_{0}^{2}\simeq2\sqrt{\pi}N_{\mathrm{p}}$.

We are then led to the important conclusion that, according to Eq. (\ref%
{Pthr}), the SPOPO threshold is roughly equal to the cw single mode
threshold $P_{0}$ divided by the number of pump modes, and can be therefore
very low. For example, if $N_{\mathrm{p}}=2\times 10^{4}$ (corresponding to $%
\tau _{\mathrm{p}}=100$ fs and a cavity length $L=2$ m) and
considering the case already discussed (a $100\,\mu$m-thick BIBO
based linear SPOPO pumped at $0.4\,\mu$m), we expect, for $T_{%
\mathrm{s}}=0.01$, a pump irradiance at threshold $P_{\text{\textrm{thr}}%
}^{\left( \text{\textrm{singly}}\right) }$ of $0.1\,\mathrm{kW}\,\mathrm{cm}^{-2}$,
and an average pump power of $16\,\text{mW}$ for a typical pump beam radius of
$70\,\mu$m.

\subsection{Numerical approach}

\label{Numdiag}

In the general case one must diagonalize numerically the $10^{5}\times10^{5}$
matrix $\mathcal{L}$. The situation can be dramatically simplified from the
computational viewpoint by noting that a scale transformation affecting the
SPOPO parameters allows diagonalization of a much smaller matrix.

Let us now consider a set of parameters defined by%
\begin{align}  \label{scalekappa a}
\beta_{1}^{\prime}&=\kappa\beta_{1}, \;\beta_{2\mathrm{p}}^{\prime}=%
\kappa^{2}\beta_{2\mathrm{p}}, \;\beta_{2\mathrm{s}}^{\prime}=\kappa^{2}%
\beta_{2\mathrm{s}}, \\
\label{scalekappa b}
N_{\mathrm{p}}^{\prime}&=\kappa^{-1}N_{\mathrm{p}}
\end{align}
with $\kappa$ a large and positive real number. Let us call $\mathcal{L}%
^{\prime}\left( m,q\right)$ the value of the matrix element with these new
parameters. The form of the matrix coefficients $\mathcal{L}\left(
m,q\right) $ when the phase mismatch coefficient $\phi_{m,q}$ has been
replaced by its approximate value (\ref{phi}) implies that :
\begin{equation}  \label{relLprimeL}
\mathcal{L}^{\prime}\left(m,q\right)= \sqrt{\kappa}\mathcal{L}\left(\kappa
m,\kappa q\right).
\end{equation}
Let us set the eigenvalue problem for $\mathcal{L}^{\prime}$:
\begin{equation}  \label{diagLprimeint}
\Lambda_{k}^{\prime}L_{k}^{\prime}\left(m\right)=\int\mathrm{d}q \mathcal{L}%
^{\prime}\left(m,q\right)L_{k}^{\prime}\left(q\right),
\end{equation}
where we added a prime to denote the new eigen-elements. Using %
\eqref{relLprimeL} and performing the change of variables $x=\kappa
m,\;y=\kappa q$, one finds that:
\begin{align}  \label{scaL}
\Lambda_{k}&=\sqrt{\kappa}\Lambda_{k}^{\prime}, \\
L_{k,m}&=L_{k}^{\prime}\left(m/\kappa\right).
\end{align}
These two relations are very useful as they allow to compute numerically
eigenvalues and eigenvectors of $\mathcal{L}$ in terms of the corresponding
ones of much smaller matrix $\mathcal{L}^{\prime}$ because, according to %
\eqref{relLprimeL}, the support of $\mathcal{L}^{\prime}$ is much reduced as
compared with that of $\mathcal{L}$. In any case the value for $\kappa$ must
be chosen adequately in the sense that the diagonalization of the toy
problem can be cast in the integral form \eqref{diagLprimeint} so as to keep
$\mathcal{L}^{\prime}$ a smooth function of $\left(m,q\right)$.

As a by-product of the demonstration an interesting prediction on the
influence of the cavity length can be drawn: consider that, given a SPOPO,
we modify its length according to $L^{\prime}=\kappa^{-1}L$. This modifies
the free spectral range as $\Omega^{\prime}=\kappa\Omega$ and the new SPOPO
parameters relate to the old ones as in \eqref{scalekappa a} and %
\eqref{scalekappa b}. Hence \eqref{diagLprimeint} leads to
\begin{equation}  \label{Newdiagonal}
\Lambda_{k}^{\prime}=\sqrt{L^{\prime}/L}\Lambda_{k}.
\end{equation}
This is the case in particular for $\Lambda_{0}$ and the new pump threshold
becomes $P_{\text{\textrm{thr}}}^{\prime}=(L/L^{\prime})P_{\text{\textrm{thr}%
}}$. Hence increasing the cavity length (and correspondingly decreasing the
repetition rate) decreases the threshold accordingly.

\section{Application of results in realistic cases}

In this Section we discuss the threshold and squeezing properties of
experimentally realizable SPOPOs. This study requires the numerical
diagonalization of the matrix $\mathcal{L}$. We have explored many different
configurations involving different pump pulse durations $\tau_{\mathrm{p}}$,
different cavity lengths $L$, different crystal thicknesses $l$ and even
different phase-matching conditions (critical and noncritical) that give
rise to different dispersion properties. We have considered both BIBO and
KNbO$_{3}$ crystals and have obtained similar results in the sense that the
analytical approach given above describes very well what is numerically
found in the region \eqref{taupmin}, no matter the particular values of the
parameters. When that condition gets violated, deviations from the analytical
result are obviously found but they affect mostly the behavior of the
eigenvectors, not so much the one of the eigenvalues. As the analytical
limit is the best also from an experimental viewpoint (there the
eigenvectors are Hermite-Gauss modes, which can be reasonably easily
produced experimentally) we consider here one case that clearly fulfills
condition \eqref{taupmin} with parameters compatible with the experimental
setup that is currently under preparation. For the sake of completeness we
also consider another one that ``slightly" violates condition \eqref{taupmin}%
. We wish to remark that these cases are representative of what we have
found in an exhaustive study. Finally, when condition \eqref{taupmin} is
more severely violated large deviations from the Hermite-Gauss case are
observed that give rise in fact to new phenomena that deserve a study on
their own and are not treated here.

The cases we discuss here correspond to collinear, degenerate type I
critical phase-matching at $0.4\mu$m pumping of a BIBO crystal \cite%
{BIBO}, obtained when the pump polarization is ordinary (parallel to the
direction $Ox$) and that of the signal is extraordinary ($o\rightarrow e+e$%
). Using Sellmeier's coefficients for BIBO we obtain that such
phase-matching occurs at an angle $\theta=151^{\circ}$ between the direction
$Oy$ and the direction of propagation of the pump (and the signal) beam, in
agreement with \cite{experimBIBO}. For this configuration we obtain
the values for the dispersion parameters reported in Table \ref{ExpParam}.
Also given in that table are the values of the free spectral range $\Omega$
and pump pulse duration $\tau_{\mathrm{p}}$ that will be used along this
section. We shall assume a pump with Gaussian spectrum and centered at the
phase-matched frequency $2\omega_0$. %\begin{table}[b!]
% \centering
%  \subtable[BIBO dispersion parameters]{
%   \begin{tabular}{c|c|c|c}
%    \hline\hline
%    $k'_{\mathrm{p}}$ (s m$^{-1}$) & $k''_{\mathrm{p}}$ (s$^2$ m$^{-1}$) & $k'_{\mathrm{s}}$ (s m$^{-1}$) & $k''_{\mathrm{s}}$ (s$^2$ m$^{-1}$)
%    \\
%    \hline
%    $6.6537\times10^{-9}$ & $4.7248\times10^{-25}$ & $6.2664\times10^{-9}$ & $1.6420\times10^{-25}$
%    \\
%    \hline
%    \end{tabular}
%  }
%  \subtable[Pump parameters]{
%    \begin{tabular}{c|c}
%     \hline\hline
%     $\Omega$ (MHz) & $\tau_{\mathrm{p}}$ (fs)
%     \\
%     \hline
%     $2\pi\times75$ & $100$
%     \\
%     \hline
%    \end{tabular}
%  }
%\caption{Dispersion parameters for BIBO crystal and pumping field.}
%\label{ExpParam}
%\end{table}
%%
%%
%\begin{align}
%k_{\mathrm{p}}^{\prime}&=6.6537\times10^{-9}\operatorname{s}\operatorname{m}^{-1},
%\nonumber
%\\
%k_{\mathrm{s}}^{\prime}&=6.2664\times10^{-9}\operatorname{s}\operatorname{m}^{-1},
%\nonumber
%\\
%k_{\mathrm{p}}^{\prime\prime}&=4.7248\times10^{-25}\operatorname{s}^{2}\operatorname{m}^{-1},
%\nonumber
%\\
%k_{\mathrm{s}}^{\prime\prime}&=1.6420\times10^{-25}\operatorname{s}^{2}\operatorname{m}^{-1}.
%\nonumber
%\end{align}
%
Finally we consider three different values of the crystal length:
\begin{align}
\text{case A}&\text{:} \;l=0.1\,\text{mm},  \notag \\
\text{case B}&\text{:} \;l=0.5\,\text{mm},  \notag \\
\text{case C}&\text{:} \;l=5\,\text{mm},  \notag
\end{align}
With all these values one can compute the phase mismatch angle $\phi$ in the
second order dispersion approximation, see Eq. \eqref{phi}, and finally the
matrix $\mathcal{L}$.
\begin{table}[htbp]
\centering
\subtable[BIBO dispersion parameters]{
   \begin{tabular}{c|c|c}
    \hline\hline
     & $k'$ (s m$^{-1}$) & $k''$ (s$^2$ m$^{-1}$)
    \\
    \hline
    pump & $6.6537\times10^{-9}$ & $4.7248\times10^{-25}$
    \\
    \hline
    signal & $6.2664\times10^{-9}$ & $1.6420\times10^{-25}$
    \\
    \hline
   \end{tabular}
  }
\subtable[Pump parameters]{
    \begin{tabular}{c|c}
     \hline\hline
     $\Omega$ (MHz) & $\tau_{\mathrm{p}}$ (fs)
     \\
     \hline
     $2\pi\times75$ & $100$
     \\
     \hline
    \end{tabular}
  }
\caption{Dispersion parameters for BIBO crystal and pumping field.}
\label{ExpParam}
\end{table}

Case A verifies well the condition \eqref{taupmin}, which reads here $\tau_{%
\mathrm{p}}>40$ fs. On the contrary cases B and C do not verify it,
which now reads $\tau_{\mathrm{p}}>200$ fs and $\tau_{\mathrm{p}}>2000$ fs, respectively.

\subsection{Case A}

\begin{figure}[b!]
\centering
\includegraphics[width=0.3\textwidth]{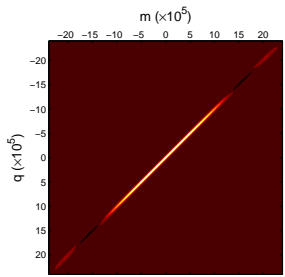}
\caption{(Color online) Case A: $\mathcal{L}$ matrix.}
\label{La}
\end{figure}

\begin{figure}[b!]
\centering
\includegraphics[width=0.4%
\textwidth]{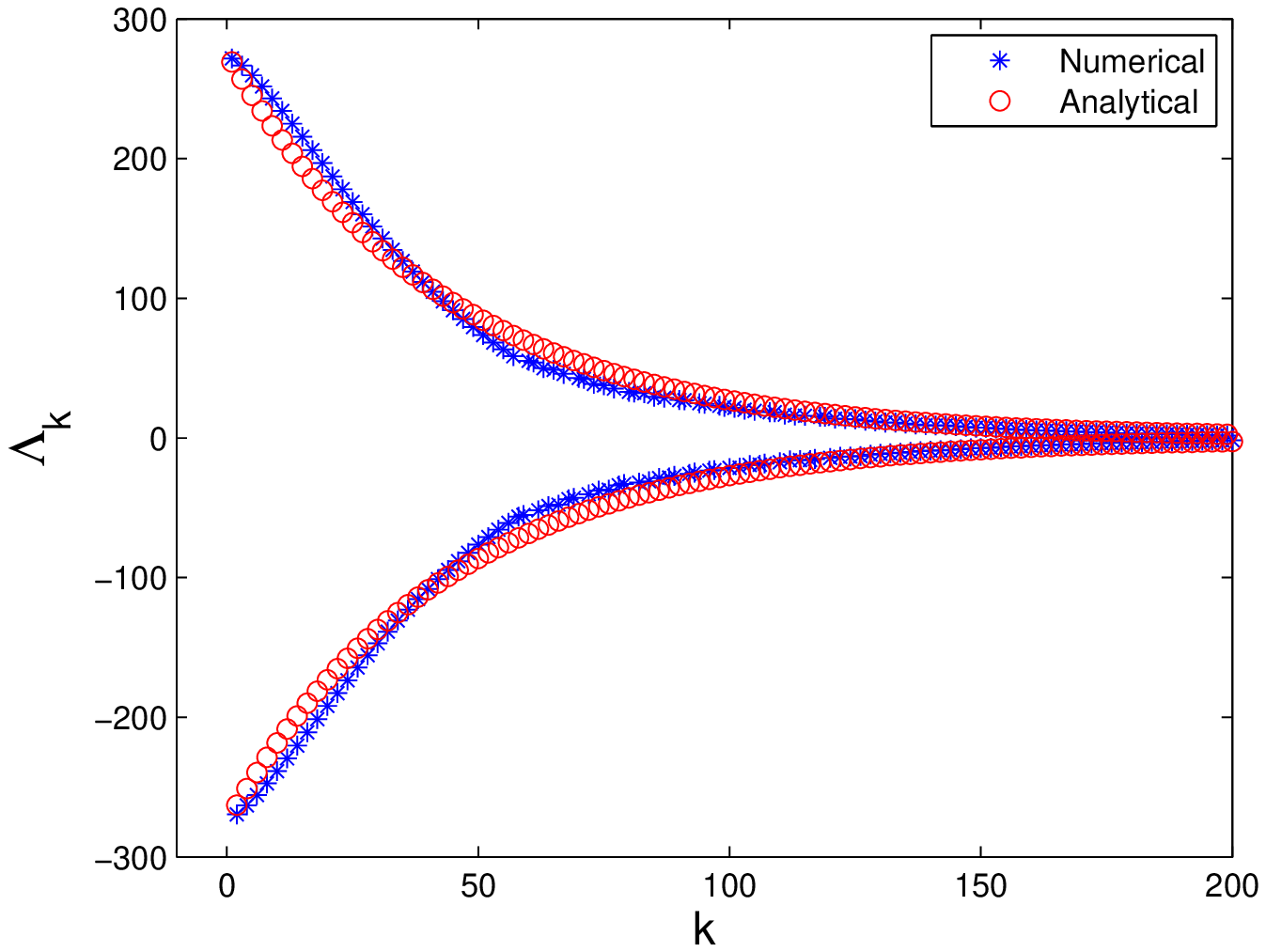}
\caption{(Color online) Case A. Spectrum of eigenvalues. Comparison between the numerical
and analytical solutions. Among about $10^{5}$ supermodes only a relatively
small part ($\sim40$) is dynamically significative.}
\label{CaseAeigenvalues}
\end{figure}

\begin{figure}[h!]
\centering
\subfigure[]{\includegraphics[width=0.2%
\textwidth]{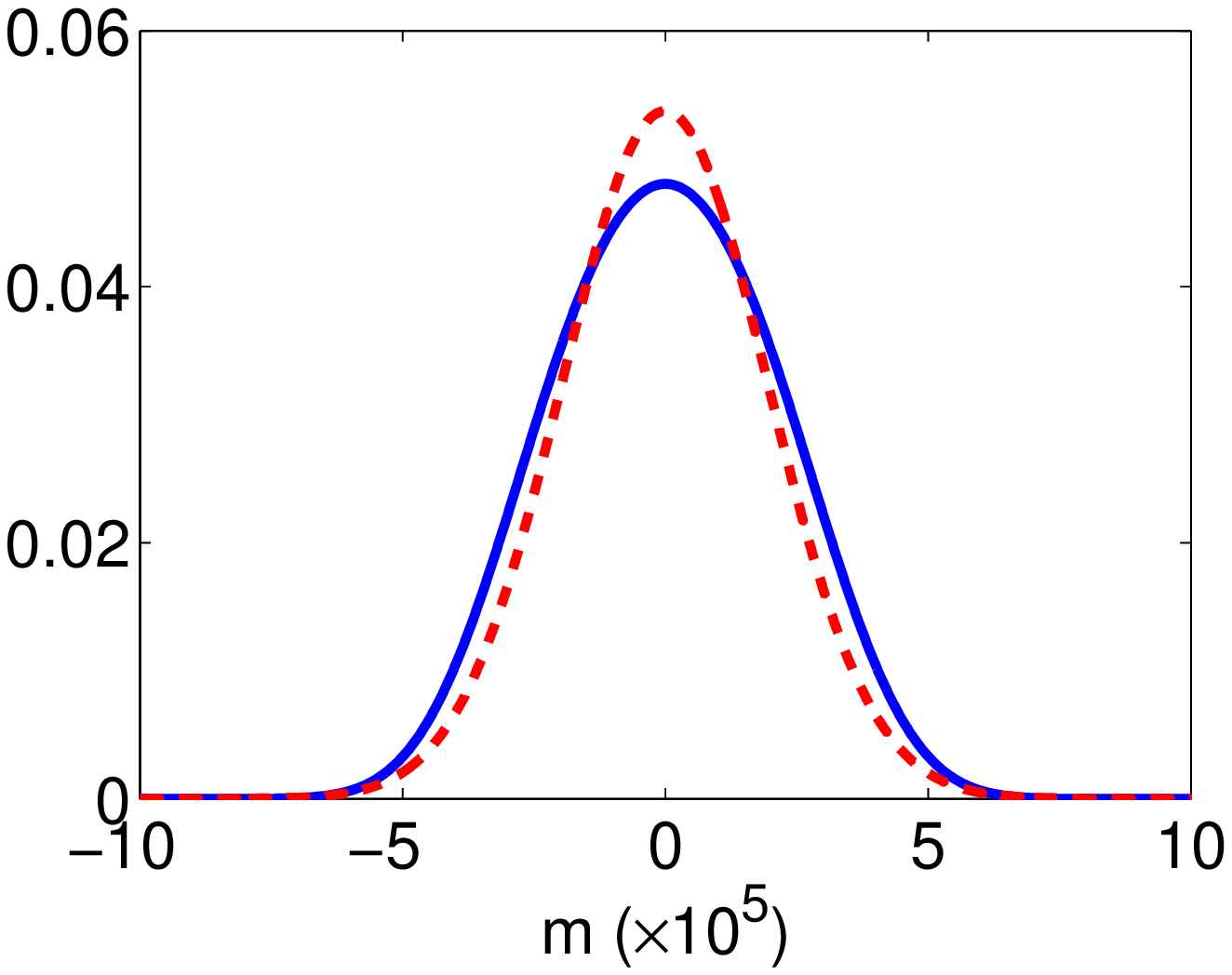}} \qquad \subfigure[]{%
\includegraphics[width=0.2\textwidth]{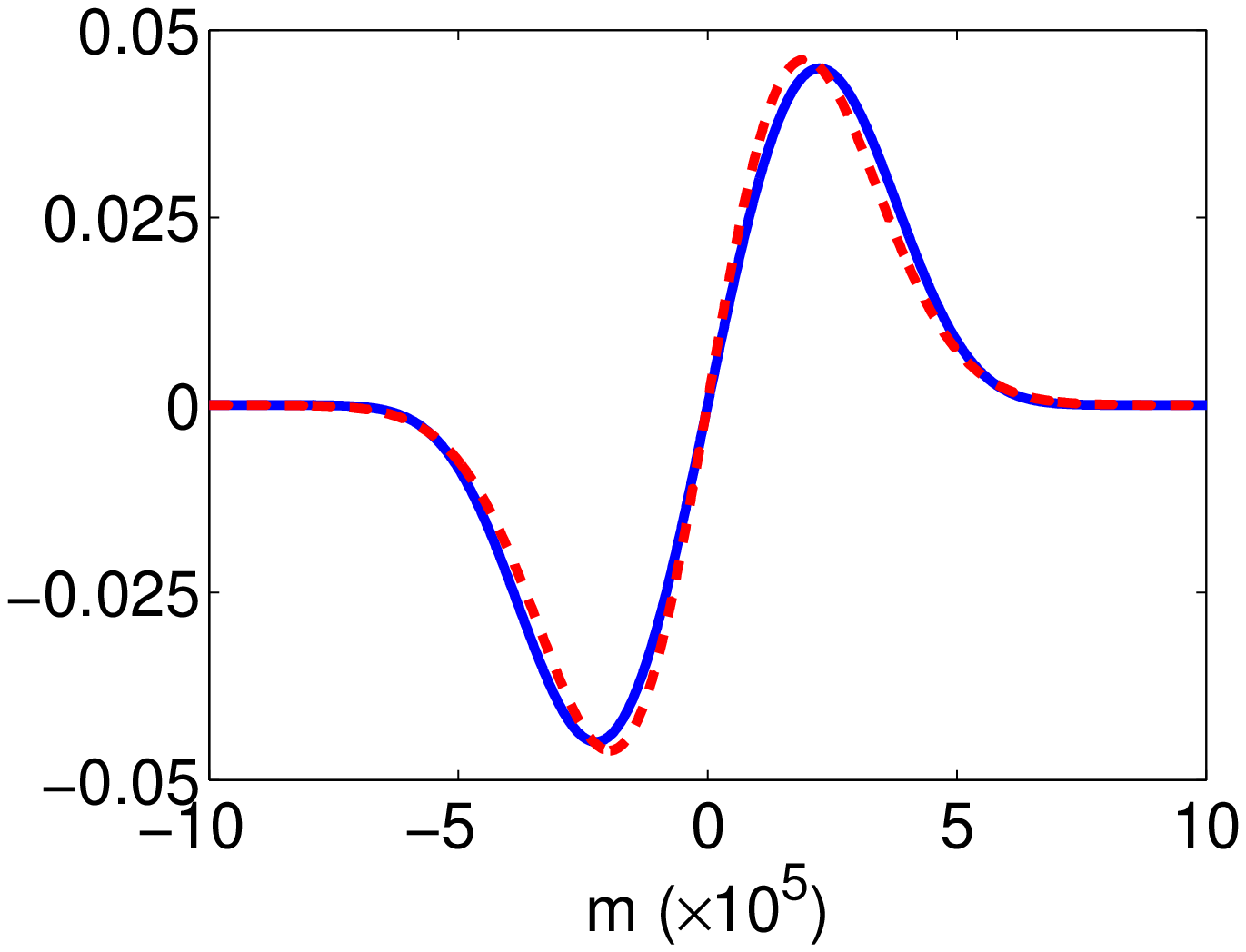}}
\qquad \subfigure[]{\includegraphics[width=0.2%
\textwidth]{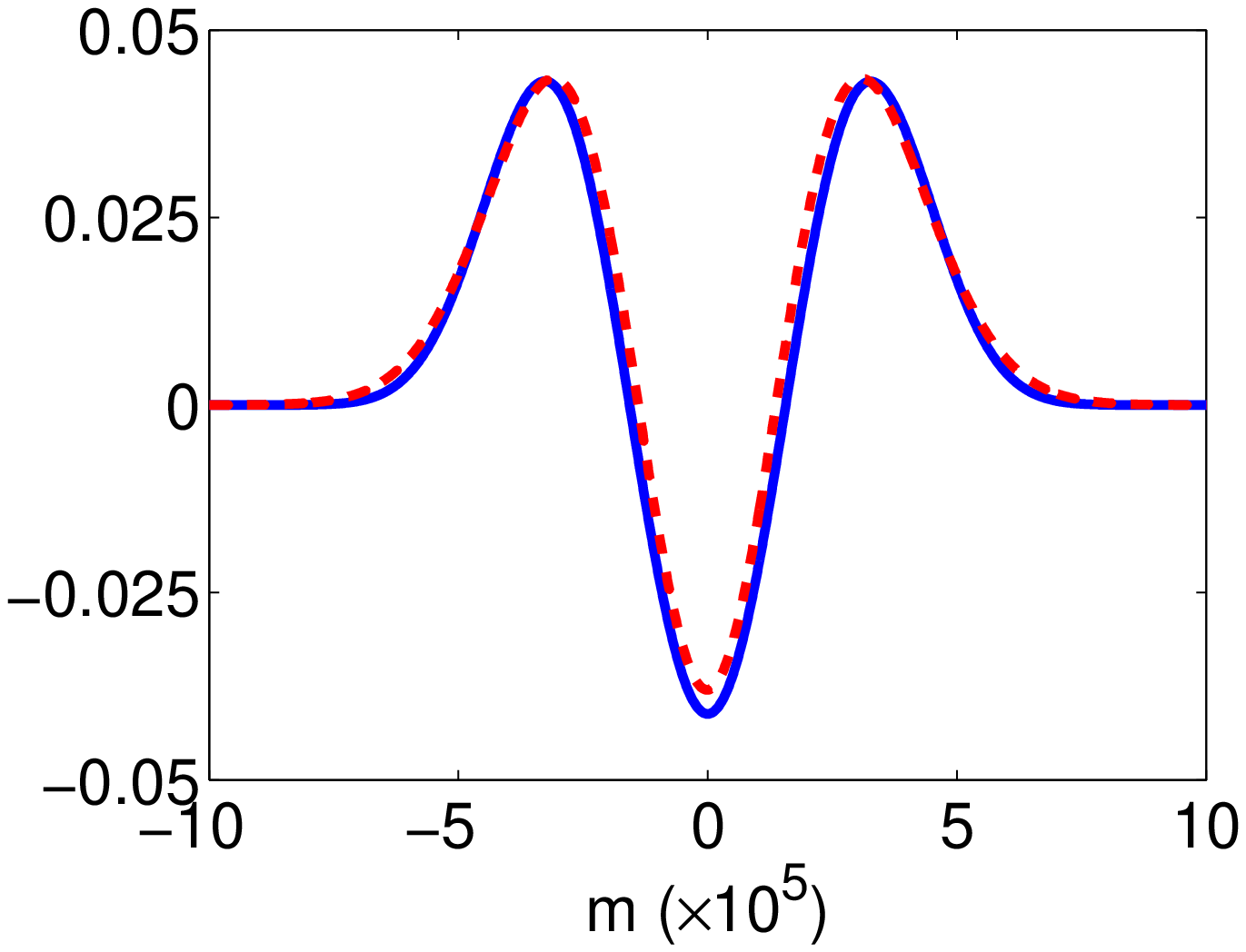}} \qquad \subfigure[]{%
\includegraphics[width=0.2\textwidth]{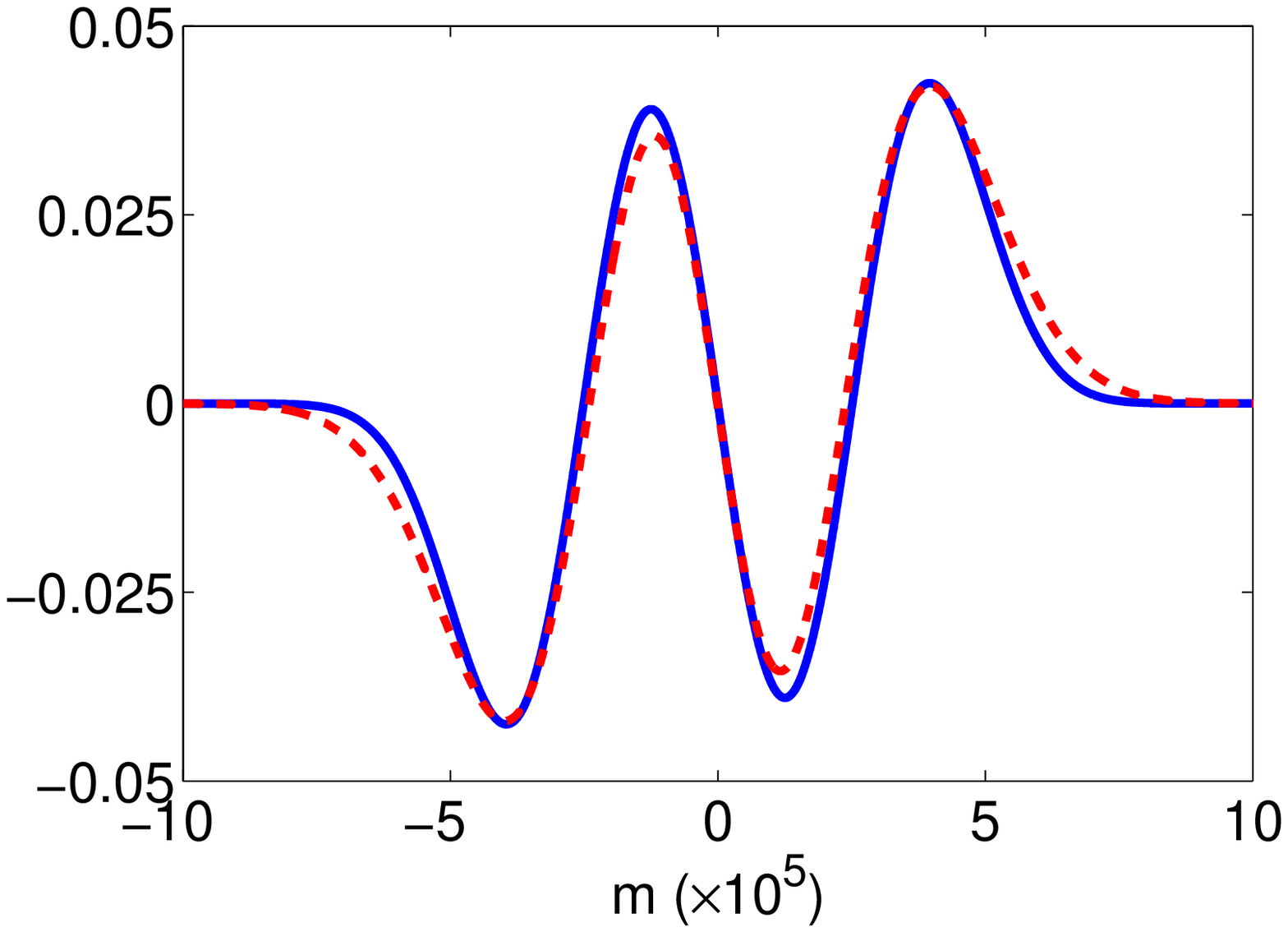}}
\caption{(Color online) Case A. Comparison between the numerical (solid blue line) and the
analytical (dotted red line) solutions of the eigenvectors associated to the
four highest $|\Lambda_k|$.}
\label{CaseAeigenvectors}
\end{figure}

In the next figure we show the matrix $\mathcal{L}$ corresponding to this
case. Fig. \ref{La} corresponds to the frequency (integer indexes)
representation, which is the actual matrix to be diagonalized. Phase
matching occurs in the lighter regions. Darker regions are highly
phase-mismatched. Results of the numerical diagonalization, obtained as
explained in Section \ref{Numdiag} by using a scale factor $\kappa=1000$,
are shown in Figs. (\ref{CaseAeigenvalues}) and (\ref{CaseAeigenvectors}).

In the singly resonant case, a threshold of $W_{\mathrm{thr}}\simeq 29\,\text{%
mW}$ is readily obtained from Eq. \eqref{Pthr} for the corresponding
eigenvalue $\Lambda_{0}\simeq 270$, by considering a transmission factor of $%
T_{\mathrm{s}}=0.01$ and a beam waist of $70\,\mu$m. This results are
in perfect agreement with the analytical value predictable by the expression
Eq. \eqref{Lam0}. For a doubly resonant cavity this result has to be
multiplied by means the correction factor that accounts for the geometry and
is reported in Table \ref{SPOPOthreshold}.

From Eqs. (\ref{VHgena}) and (\ref{VHgenb}) we can calculate also the noise
reduction corresponding to the first four eigenvectors shown in Fig. 4 for a
zero noise frequency. Evidently we are assuming to be able to master the
spectral shape of the local oscillator in order to exactly match it to the
supermode whose noise variance spectrum is to be measured.
Nevertheless, the optimization of mode-matching between the local oscillator
and a specific supermode can result a difficult task even when pulse shaping
techniques are used. Let's consider, then, the case where the best we can do
is to deal with a local oscillator shaped as a Gauss-Hermite polynomial $%
e_{k,m}=\pi^{-1/4}N_{\mathrm{L}}^{-1/2}\mathrm{e}^{-\frac{1}{2}\left(m/N_{%
\mathrm{L}}\right)^{2}} \mathrm{e}^{i\phi_{\mathrm{L}}}H_{k}\left(m/N_{%
\mathrm{L}}\right)$, where $N_{\mathrm{L}}$ is the number of longitudinal
modes of the local oscillator comb. In such situation the variances have to
be evaluated using the general expression Eq. \eqref{Squeezingspectrum},
where the noise variance spectrum is given by the sum of all the supermodes
noise variance spectra weighted by the mode matching parameters $d_k$, which
describe how well each supermode projects on the local oscillator field. In
Table \ref{Vars09A} we compare the degree of squeezing measured in the
situation of perfect mode-matching and the situation where the best local
oscillator is Gauss-Hermite function of adjustable spectral width, for a
pumping power $20\%$ below threshold (i.e. $r=0.9$).
\begin{table}[t!]
\centering
\begin{tabular}{c|c|c|c|c}
\hline\hline
$V_{k}$ (dB) & $k=0$ & $k=1$ & $k=2$ & $k=3$ \\ \hline
perfect & $-25.6$ & $-24.9$ & $-24.1$ & $-23.3$ \\ \hline
G-H & $-25.5$ & $-24.8$ & $-23.8$ & $-22.6$ \\ \hline
\end{tabular}%
\caption{Case A. Comparison between the noise variances evaluated, at $%
\protect\omega=0$ and $r=0.9$, in the case of perfect mode matching of the
LO with the supermodes corresponding to $k=0,1,2,3$ and the case where the
LO is the Gauss-Hermite (G-H) function described by the spectral amplitudes $%
e_{k,m}$ and $N_{\mathrm{L}}=2.2\times10^{5}$. }
\label{Vars09A}
\end{table}
In the latter case, the minimum noise is obtained around $N_{\mathrm{%
L}}\simeq2.2\times10^5$. Such comparison evidences the fact that the
differences between the two situations are small. Hence, the exact knowledge
of the supermodes shape is not necessary and a good degree of mode matching
can be obtained simply controlling the spectral width of a Gauss-Hermite
local oscillator. Evidently this circumstance is verified as far as the
condition for the Gaussian approximation of the coupling matrix is respected.%
However, the number of supermodes that present marked quantum
characteristics results to be greater than four. By considering, in a
qualitative way, that $-5\,\text{dB}$ is a still significative degree of
squeezing, we found that all the supermodes corresponding to the first $45$
higher values of $|\Lambda_k|$ have variances smaller than the considered
bound. This is an important result since it proves that SPOPOs are
multi-mode sources of non-classical light.

\subsection{Case B}

\begin{figure}[h!]
\centering
\includegraphics[width=0.3\textwidth]{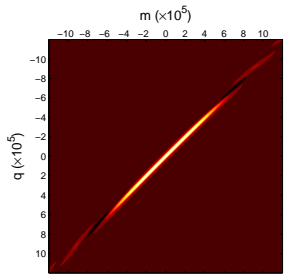}
\caption{(Color online) Case B: $\mathcal{L}$ matrix.}
\label{LcaseBa}
\end{figure}

In the next figure we show the matrix $\mathcal{L}$ corresponding to this
case. Fig. \ref{LcaseBa} corresponds to the frequency (integer indexes)
representation, which is the actual matrix to be diagonalized. Phase
matching occurs in the lighter regions, while darker regions are highly
phase-mismatched.

Results of the numerical diagonalization, obtained as explained in Section %
\ref{Numdiag} by using a scale factor $\kappa=1000$, are shown in Figs. \ref%
{Case B eigenvalues} and \ref{Case B eigenvectors}.
\begin{figure}[t!]
\centering
\includegraphics[width=0.4%
\textwidth]{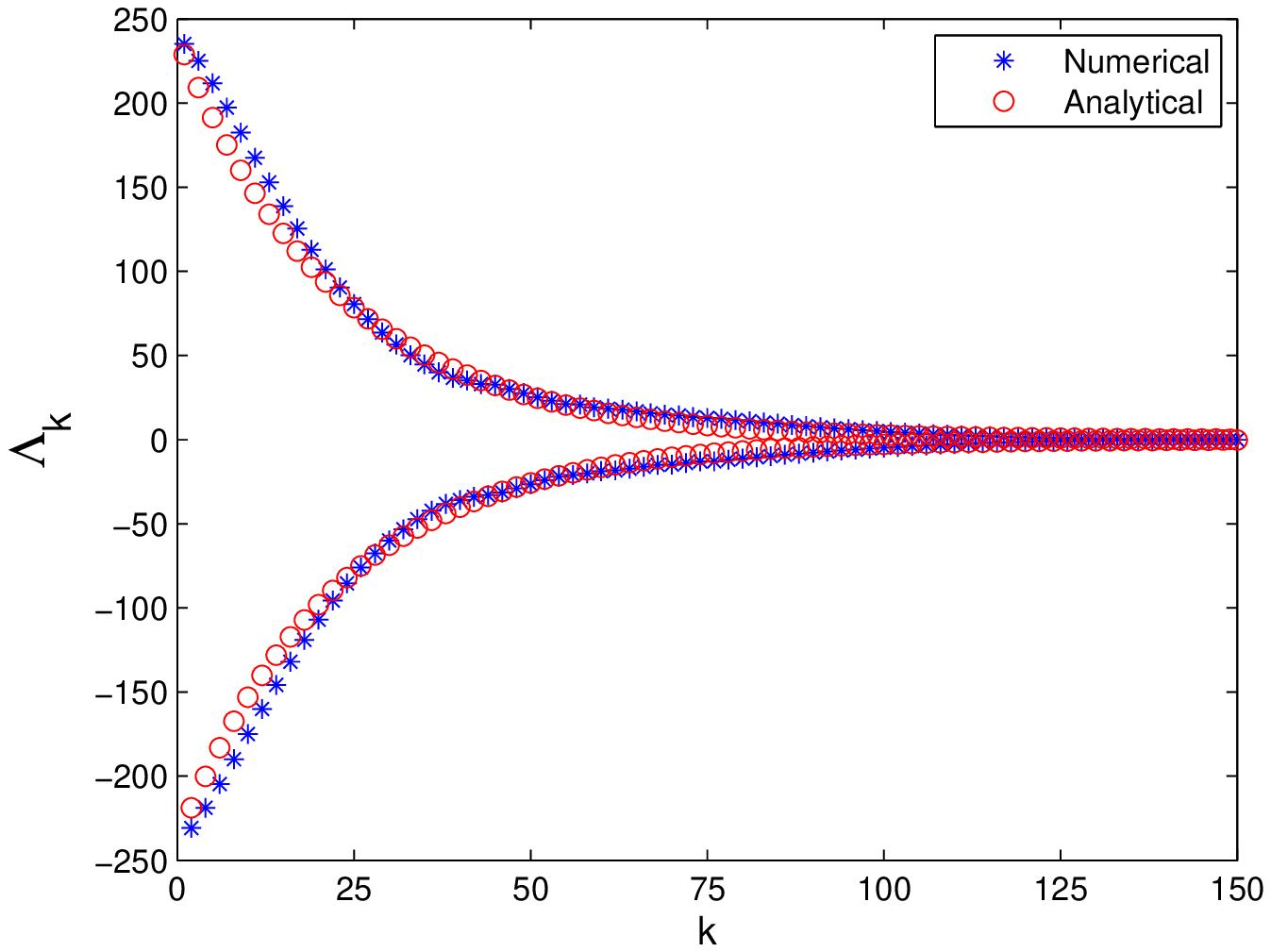}
\caption{(Color online) Case B. Spectrum of eigenvalues. Comparison between the numerical
and analytical solutions. Among about $10^{5}$ supermodes only a relatively
small part ($\sim23$) is dynamically significative.}
\label{Case B eigenvalues}
\end{figure}
\begin{figure}[t!]
\centering
\subfigure[]{\includegraphics[width=0.2%
\textwidth]{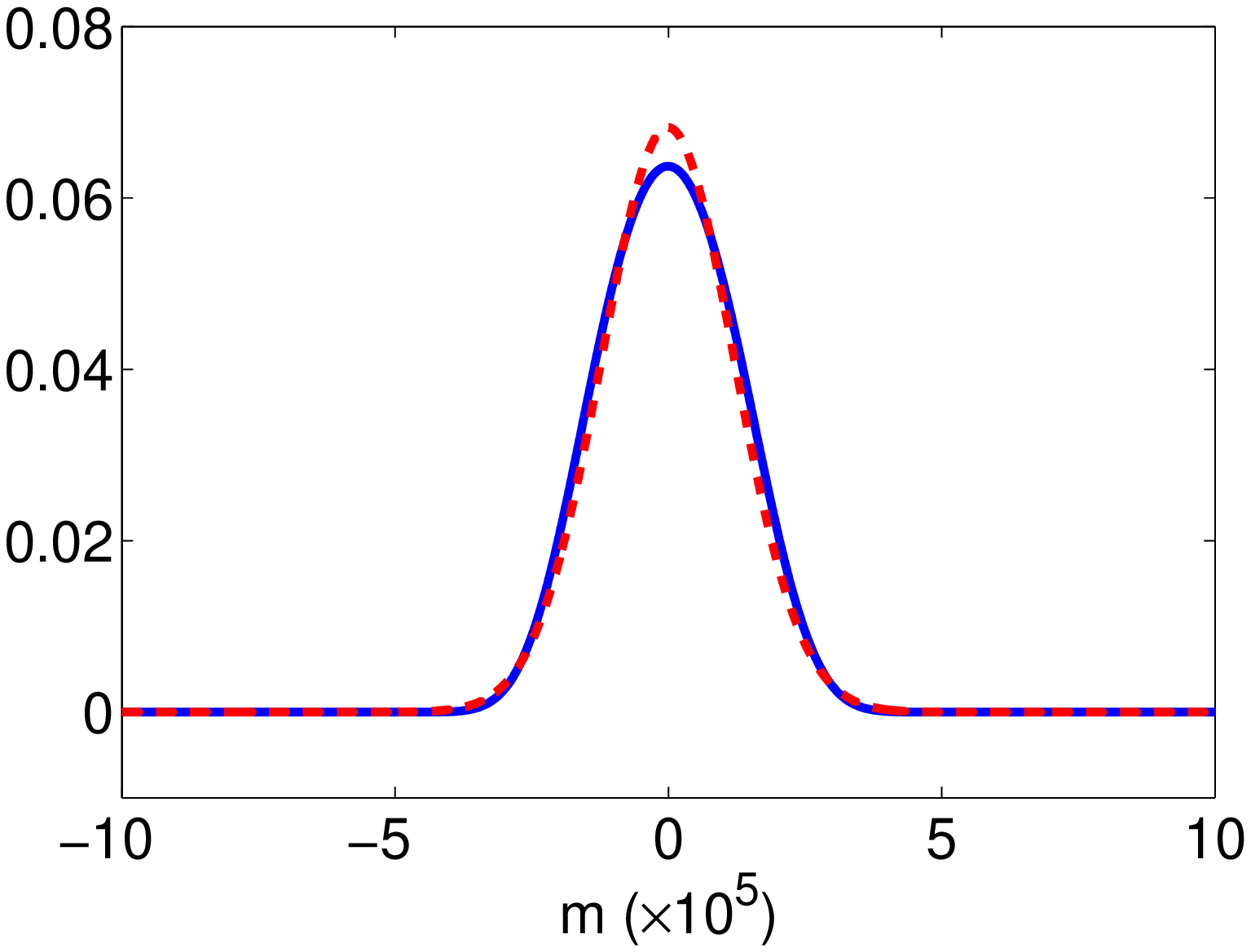}} \qquad %
\subfigure[]{\includegraphics[width=0.2%
\textwidth]{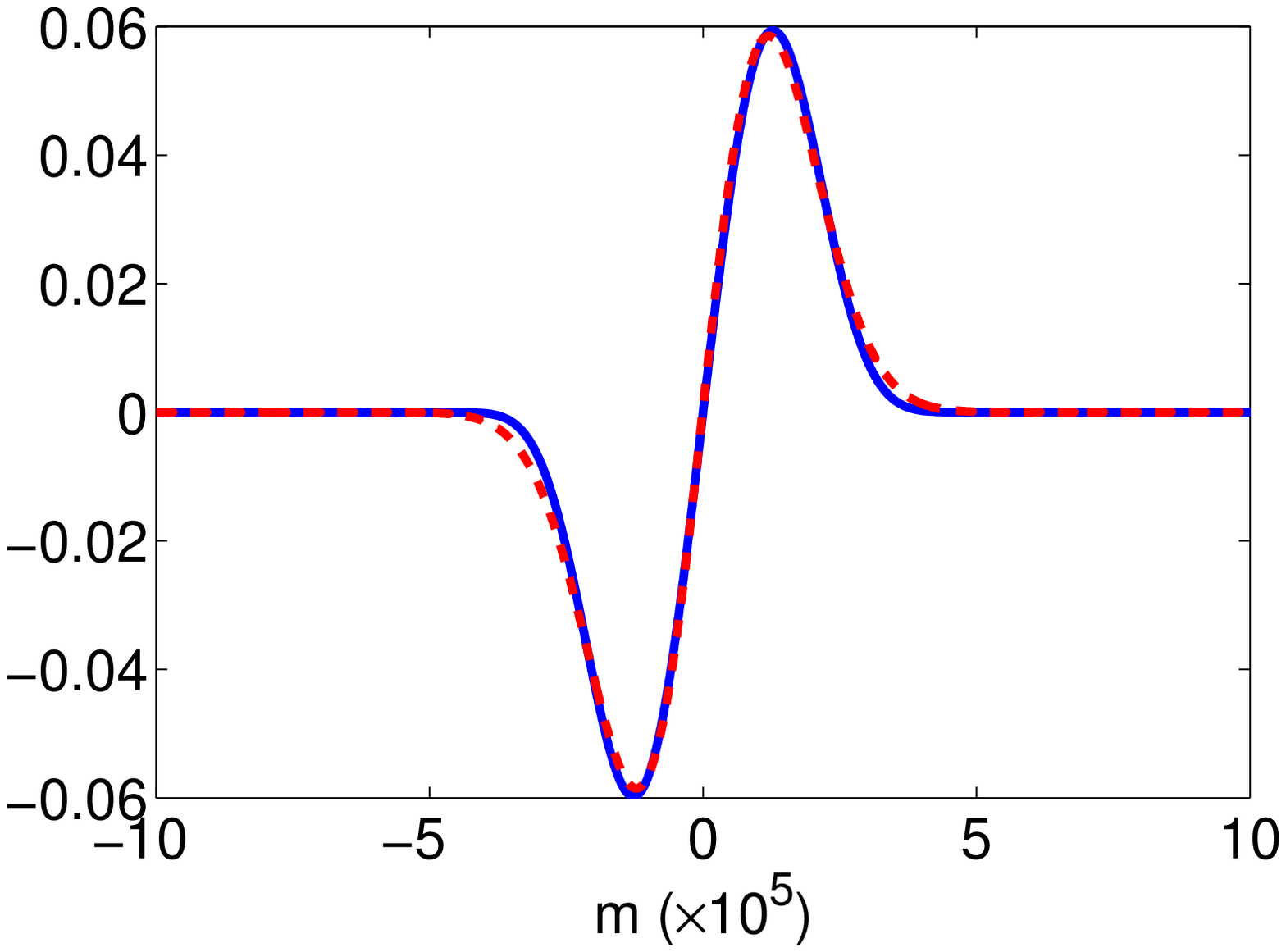}} \qquad %
\subfigure[]{\includegraphics[width=0.2%
\textwidth]{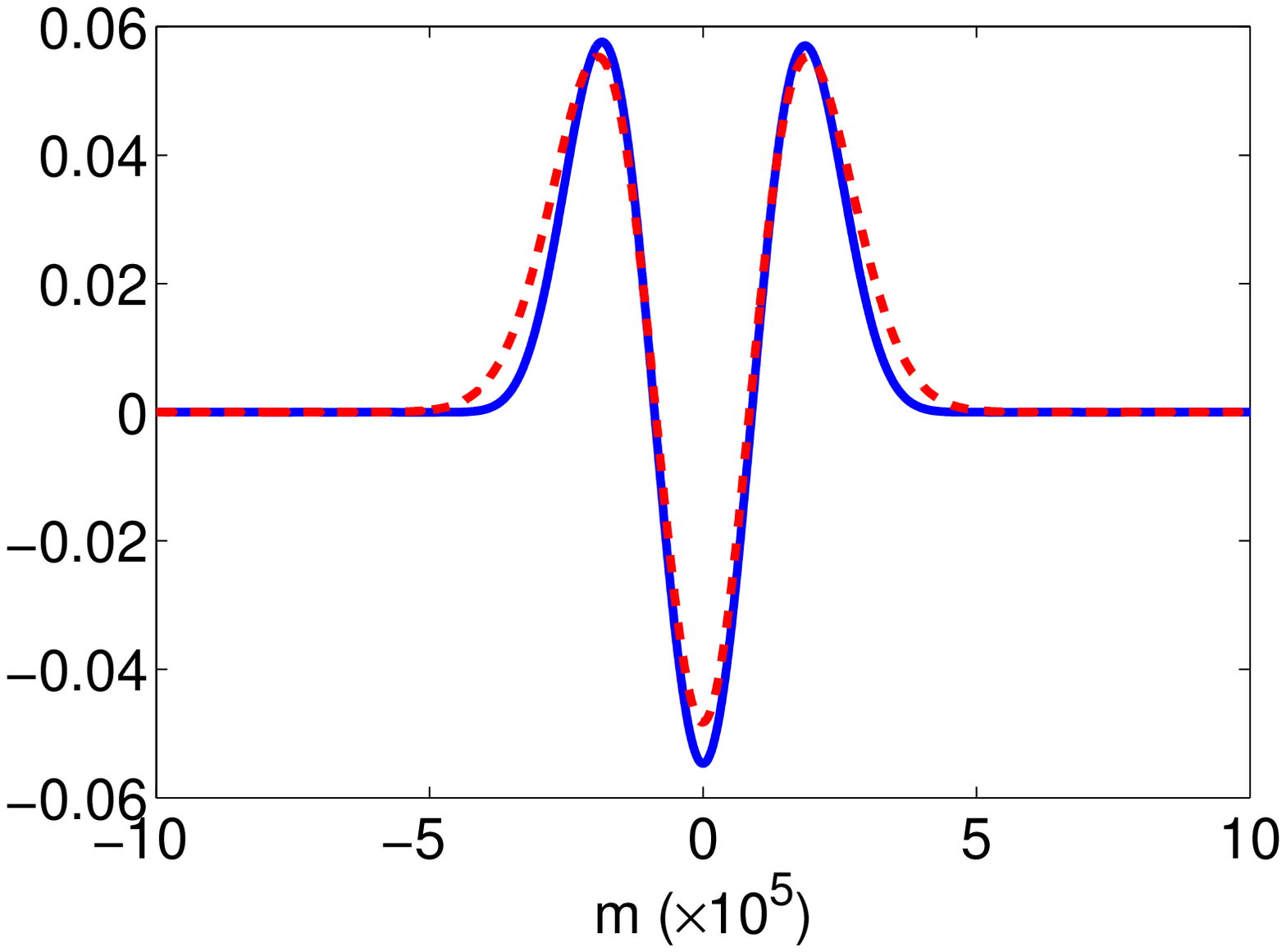}} \qquad %
\subfigure[]{\includegraphics[width=0.2%
\textwidth]{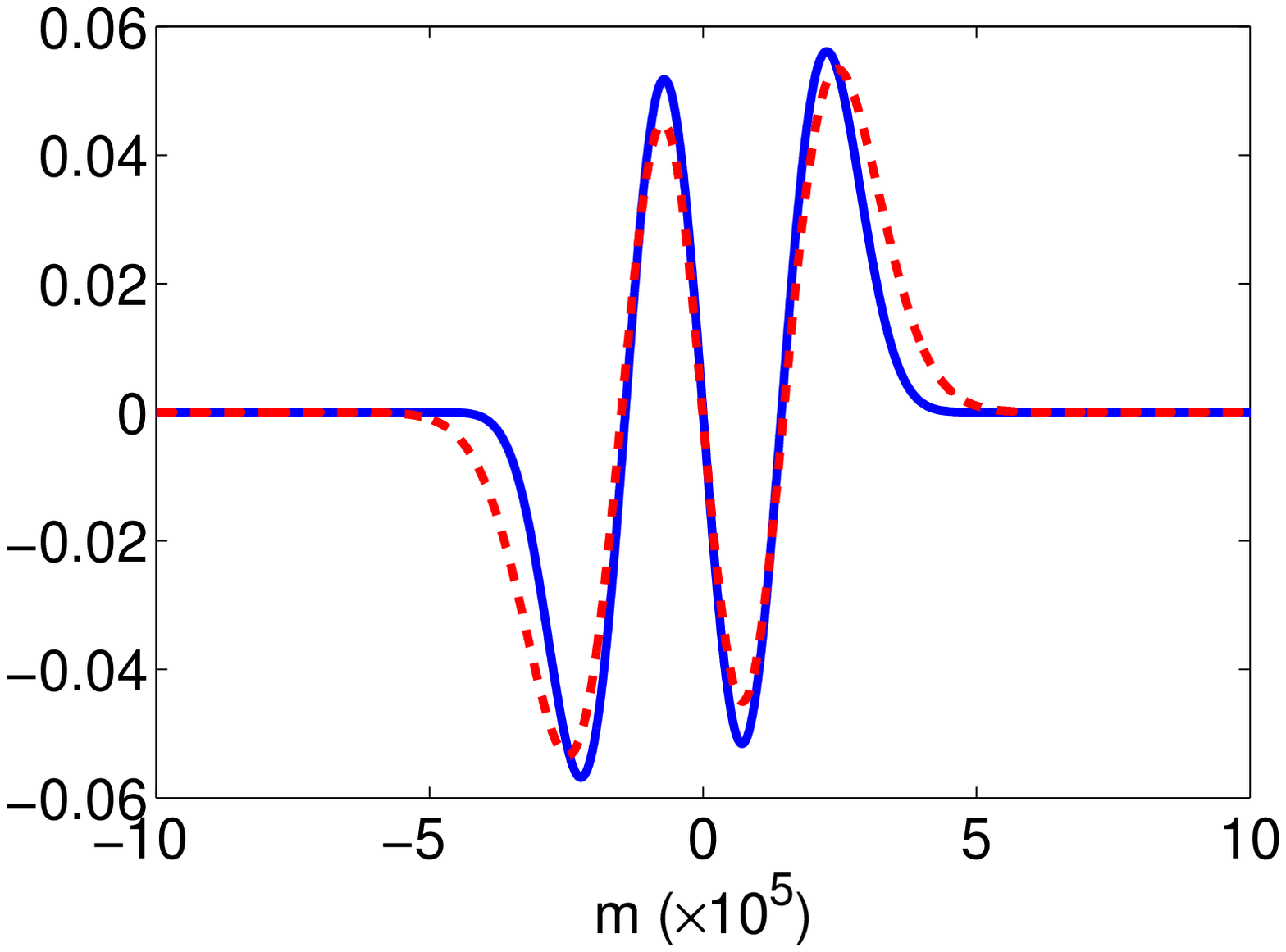}}
\caption{(Color online) Case B. Comparison between the numerical (solid blue line) and the
analytical (dotted red line) solutions of the eigenvectors associated to the
four highest $|\Lambda_k|$.}
\label{Case B eigenvectors}
\end{figure}

In the singly resonant case, a threshold of $W_{\mathrm{thr}}\simeq 1.5\,\text{%
mW}$ is readily obtained from Eq. \eqref{Pthr} for the corresponding
eigenvalue $\Lambda_{0}\simeq 235$, by considering a transmission of $T_{%
\mathrm{s}}=0.01$ and a beam waist of $70\,\mu$m. The threshold
obtained from the analytical solution is about $1.6\,\text{mW}$, which
is not too much different from the exact solution. Hence, even if the
experimental situation considered here does not strictly verify the
conditions for Gaussian approximation, we find still a good agreement
between the numerical and the analytical predictions. A qualitative
statement about the good agreement between the numerical and analytical
solutions can be grounded also from the comparison between the eigenvalues
shown in Fig. \ref{Case B eigenvectors}.

Assuming perfect mode matching, from Eqs. \eqref{VHgena} and %
\eqref{VHgenb} we can calculate also the noise reduction corresponding to
the first four eigenvectors shown in Fig. \ref{Case B eigenvectors} at the
carrying frequency and $20\%$ below threshold (i.e. $r=0.9$) and
compare it to the general case where the local oscillator is described by
the Gauss-Hermite spectral amplitudes $e_{k,m}$. In this case, the detection
is optimized for a spectral width of about $N_{\mathrm{L}}\simeq1.2%
\times10^{5}$. The results are reported in Table \ref{Vars09B}.
\begin{table}[t!]
\centering
\begin{tabular}{c|c|c|c|c}
\hline\hline
$V_{k}$ (dB) & $k=0$ & $k=1$ & $k=2$ & $k=3$ \\ \hline
perfect & $-25.6$ & $-24.1$ & $-22.6$ & $-21.0$ \\ \hline
G-H & $-23.9$ & $-20.8$ & $-17.9$ & $-15.0$ \\ \hline
\end{tabular}%
\caption{Case B. Comparison between the noise variances evaluated, at $%
\protect\omega=0$ and $r=0.9$, in the case of perfect mode matching of the
LO with the supermodes corresponding to $k=0,1,2,3$ and the case where the
LO is the Gauss-Hermite (G-H) function described by the spectral amplitudes $%
e_{k,m}$ and $N_{\mathrm{L}}=1.2\times10^{5}$. }
\label{Vars09B}
\end{table}
In this case, the amount of squeezing detected using a Gauss-Hermite
local oscillator, despite of the optimization of its spectral width, is not
as much as the perfect case, even if still significative. This result is a
consequence of the fact that the situation now considered is slightly
violating the bounds for the Gaussian approximation and, hence, the
supermodes have spectral amplitudes that are no more characterized by
Gauss-Hermite functions. Nevertheless, even if Gaussian approximation is
not perfect, its prediction capability is still relevant.

Since we are interested to SOPOs as multi-mode sources for
non-classical light, let's consider the same qualitative argument we
considered in the previous section. In this case about 23 supermodes present
a degree of squeezing greater -5 dB. Despite the fact that, with respect to
the 0.1mm-thick crystal, the number of supermodes that characterize the
SPOPOs output is smaller, it can still be considered highly multi-mode.

\subsection{Case C}

\begin{figure}[h!]
\centering
\includegraphics[width=0.3%
\textwidth]{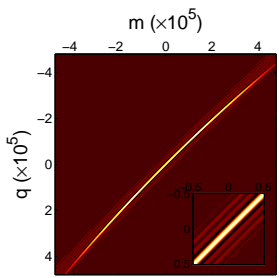}
\caption{(Color online) Case C: $\mathcal{L}$ matrix. The inset is the magnification of the
matrix around the phase-matched frequency corresponding to $m=0,\,q=0$.}
\label{LcaseC}
\end{figure}

The last case we consider corresponds to a configuration that is strongly
non-Gaussian. This can be directly observed by comparing the matrix $%
\mathcal{L}$ obtained in this case and reported in Fig. \ref{LcaseC} with
the matrices for cases A and B.
\begin{figure}[t!]
\centering
\includegraphics[width=0.4\textwidth]{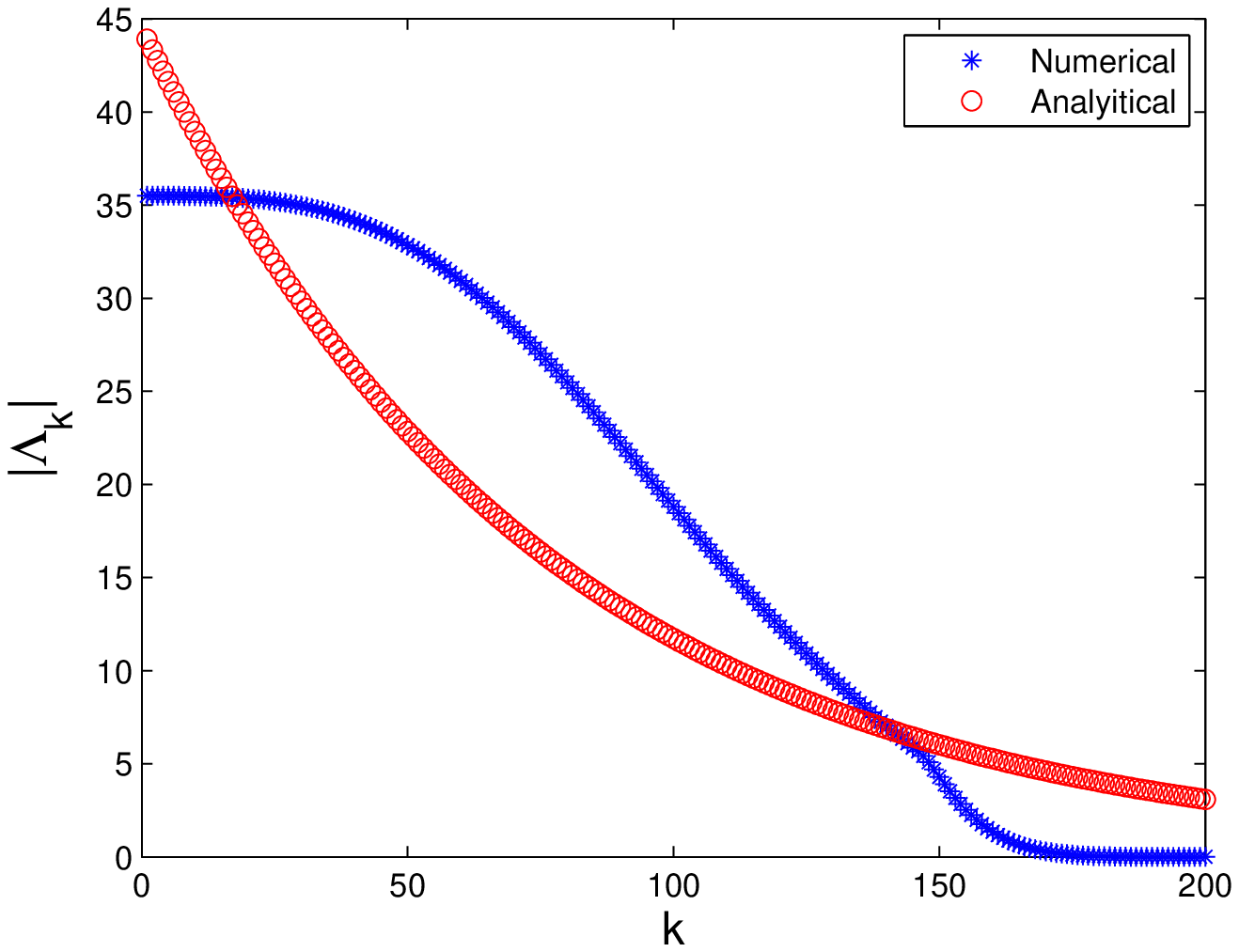}
\caption{(Color online) Case B. Spectrum of eigenvalues. Comparison between the numerical
and analytical solutions. Among about $10^{5}$ supermodes only a relatively
small part ($\sim125$) is dynamically significative.}
\label{CaseCeigenvalues}
\end{figure}
\begin{figure}[h!]
\centering
\subfigure[]{\includegraphics[width=0.2%
\textwidth]{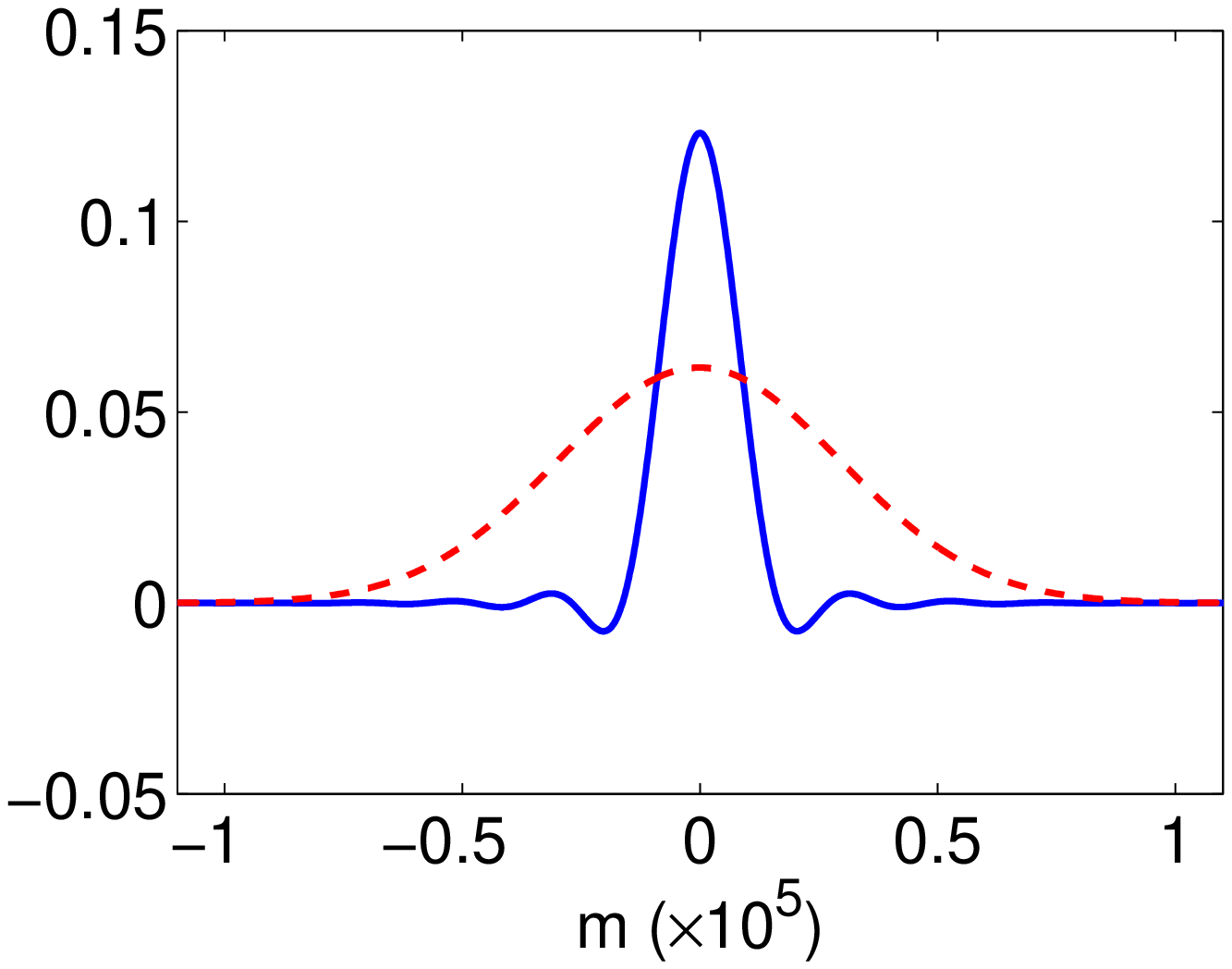}} \qquad %
\subfigure[]{\includegraphics[width=0.2%
\textwidth]{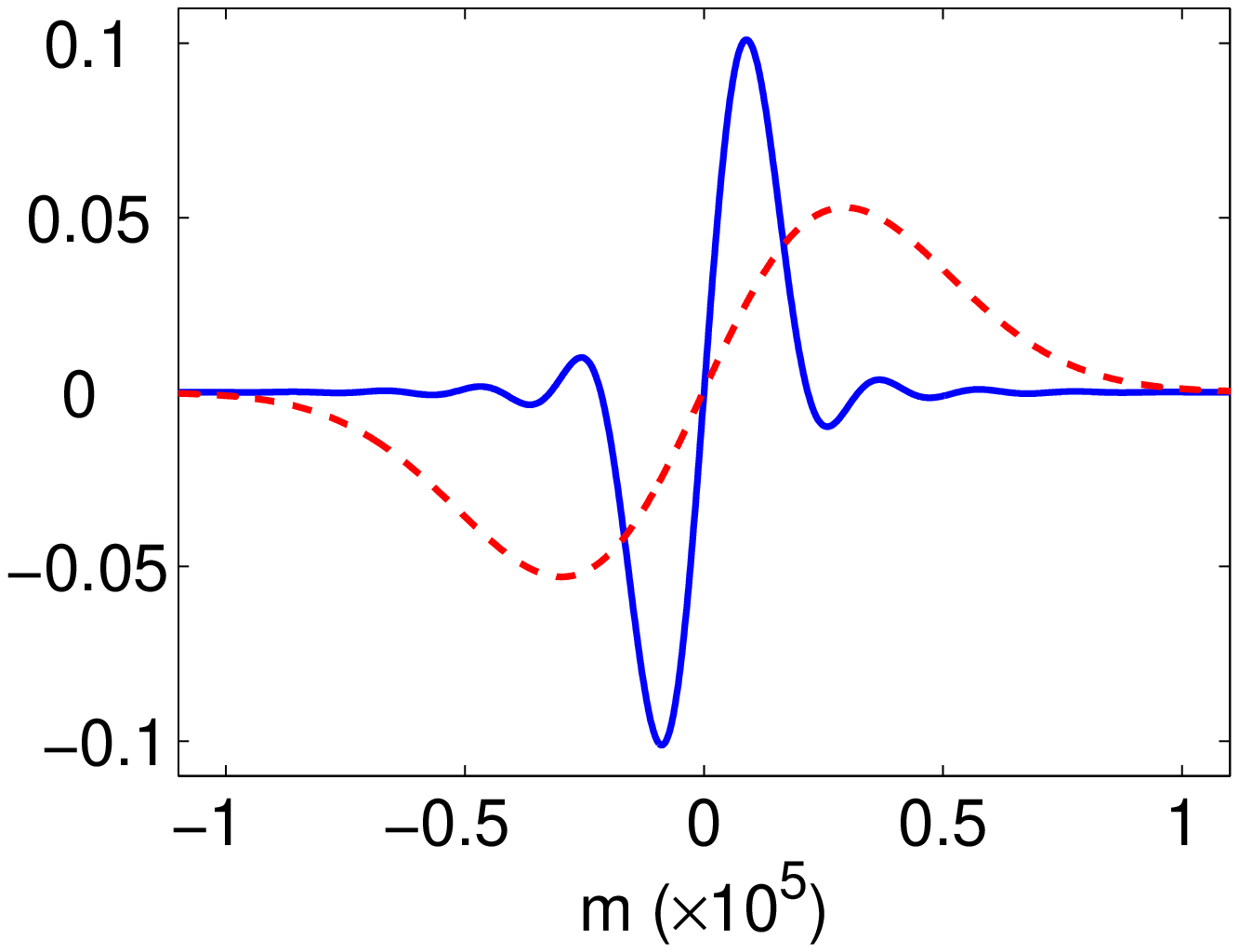}} \qquad %
\subfigure[]{\includegraphics[width=0.2%
\textwidth]{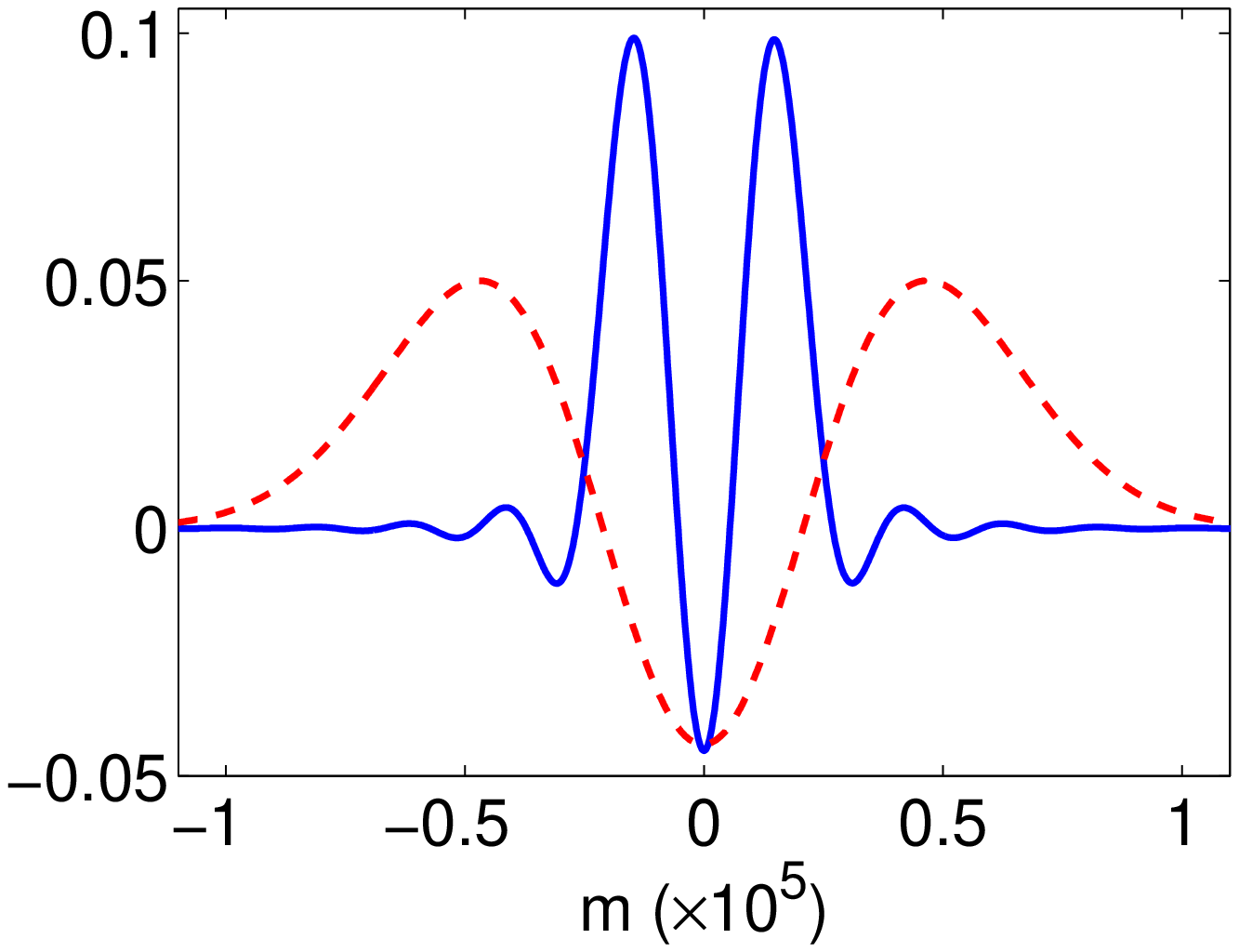}} \qquad %
\subfigure[]{\includegraphics[width=0.2%
\textwidth]{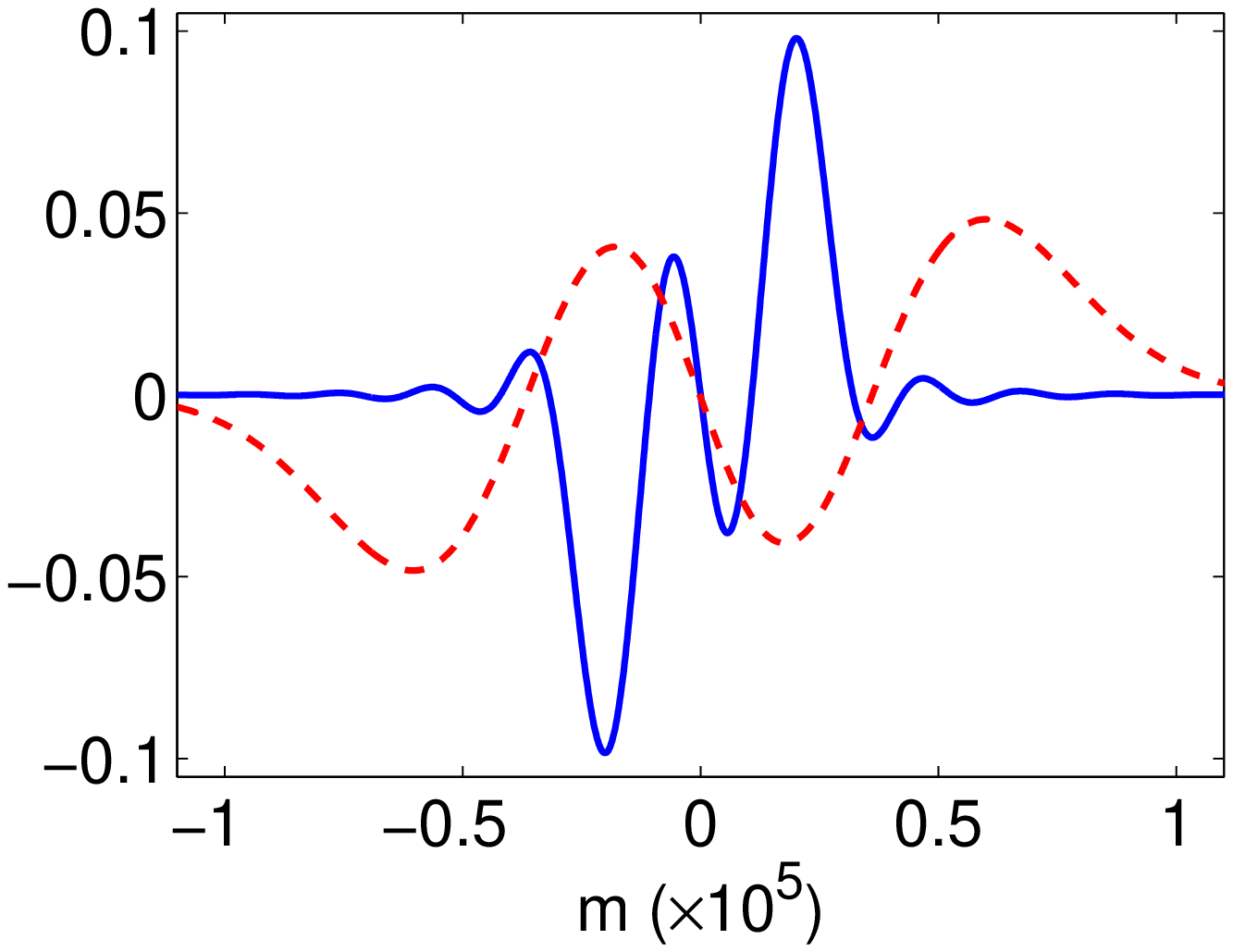}}
\caption{(Color online) Case C. Comparison between the numerical (solid blue line) and the
analytical (dotted red line) solutions of the eigenvectors associated to the
four highest $|\Lambda_k|$.}
\label{CaseCeigenvectors}
\end{figure}
In fact, since the pump pulse duration $\tau_{\mathrm{p}}$ is much smaller
than the temporal walk-off $|k^{\prime }_{\mathrm{p}}-k^{\prime }_{\mathrm{s}%
}|l$, in the space $\{m,q\}$, the pump bandwidth $N_{\mathrm{p}}$ is larger
than the phase-matching bandwidth $N_{1}$ (see Fig. \ref{Legenda}).
Consequently, the pump selects not only the principal peak of the ``sinc"
function corresponding to the phase-matching matrix $f_{m,q}$, but also
several secondary maxima. This has an important consequence for what
concerns the eigenvectors and eigenvalues of $\mathcal{L}$, obtained as
explained in Section \ref{Numdiag} by using a scale factor $\kappa=200$,
that we report in Figs. \ref{CaseCeigenvalues} and \ref{CaseCeigenvectors}.

In Fig. \ref{CaseCeigenvalues}, the spectrum of the eigenvalues obtained
from the analytical solution \eqref{LamGauss} (red circles) shows a great
discrepancy with the eigenvalues obtained from numerical diagonalization of $%
\mathcal{L}$, as expected. In particular, the part of spectrum,
corresponding about to the first $50$ eigenvalues, flatten around the
critical value $\Lambda_0\simeq36$, while the Gaussian approximation
predicts always a geometric progression-like behavior with a critical
eigenvalue $\Lambda^{\mathrm{Gauss}}_{0}\simeq44$. In the same experimental
configuration as previous cases (singly resonant cavity, transmission at
signal frequencies of $T_{\mathrm{s}}=0.01$ and beam waist of $70\,\mu$m), a threshold of $W_{\mathrm{thr}}\simeq 0.67\,\text{mW}$ can be obtained
from Eq. \eqref{Pthr}.

The result of a threshold higher than the one expected in the Gaussian
approximation has a physical explanation. Since, for the time durations
involved, the process of parametric down conversion can be considered almost
instantaneous, for a mode-locked pumping field the peak power necessary to
reach the oscillation threshold is the result of the coherent contribution
of all its modes. This circumstance is formally expressed by the fact that
the analytical expression for $\Lambda_0$ (see Eq. \eqref{LamGauss}) depends
on the number of modes in the pump pulse $N_{\mathrm{p}}$. But, beyond the
Gaussian limit, the fact that $N_{\mathrm{p}}\gg N_1$ implies that not all
the $N_{\mathrm{p}}$ pump modes are equally phase-matched and, then, not all
can optimally transfer energy towards the signal modes. The same phenomenon
can be understood even in the temporal domain. In fact the quantity $%
|k^{\prime }_{\mathrm{p}}-k^{\prime }_{\mathrm{s}}|l$ corresponds to the
temporal walk-off accumulated by the pump and signal pulses through a
passage in the nonlinear crystal. When, in a non-Gaussian regime, the
condition \eqref{taupmin} is violated, the walk-off between pump and signal
is bigger than the pump width $\tau_{\mathrm{p}}$ and the two field cannot
optimally exchange energy all along the crystal length thus increasing the
instantaneous peak power necessary to reach the oscillation.

On the other hand, in Fig. \ref{CaseCeigenvectors}, the eigenvectors
retrieved in the Gaussian approximation \eqref{LkGauss} do no more fit the
numerical solutions, as expected. In fact, even if they still preserve a
shape similar to Gauss-Hermite functions, they result to be shorter in the
domain of frequencies and are affected by a small modulation of the spectral
amplitude.

As the previous cases, we consider the quantum properties of the supermodes
and compare the situation of perfect mode matching of the LO with that of a
Gauss-Hermite LO. In this latter case the detection is optimized for a
spectral width of about $N_{\mathrm{L}}\simeq0.08\times10^{5}$. The results
are reported in Table \ref{Vars09C} for the eigenvectors corresponding to
the first four biggest $|\Lambda_{k}|$ .
\begin{table}[t!]
\centering
\begin{tabular}{c|c|c|c|c}
\hline\hline
$V_{k}$ (dB) & $k=0$ & $k=1$ & $k=2$ & $k=3$ \\ \hline
perfect & $-25.58$ & $-25.58$ & $-25.57$ & $-25.57$ \\ \hline
G-H & $-25.27$ & $-24.28$ & $-22.51$ & $-20.22$ \\ \hline
\end{tabular}%
\caption{Case C. Comparison between the noise variances evaluated, at $%
\protect\omega=0$ and $r=0.9$, in the case of perfect mode matching of the
LO with the supermodes corresponding to $k=0,1,2,3$ and the case where the
LO is the Gauss-Hermite (G-H) function described by the spectral amplitudes $%
e_{k,m}$ and $N_{\mathrm{L}}=0.08\times10^{5}$. }
\label{Vars09C}
\end{table}
The fact that they are close to degeneracy (see Fig. \ref{CaseCeigenvalues})
is reflected in an almost equal reduction of noise variances below the
standard quantum limit. Despite the differences reported between the
variances evaluated both by means of a perfectly mode matched and a
Gauss-Hermite LO, a still significative degree of squeezing can be detected
in realistic situations thus suggesting that the shaping of the LO is not a
critical issue for the experimental configuration considered in this
section. Actually, there is a larger set of supermodes the variances of
which are all close to the value of the critical one. In particular, there
are about $30$ supermodes that have variances comprised in $1$ dB, between $%
-24.6$ dB and $-25.6$ dB. Furthermore, by considering the number of supermodes
that present a noise reduction bigger than $-5$ dB, one discovers that, this
time, their number amounts to about $125$.

\subsection{Discussion on the influence of the crystal length}

We have seen that even if the cases A and B do not verify at the same
time the condition \eqref{taupmin}, the analytical solution obtained in the
Gaussian approximation works quite well in both cases. Nevertheless, in
spite of the fact that this condition gives an approximately good idea of
the reliability of Gaussian approximation, it is interesting to study the
passage from a perfectly Gaussian case to a non-Gaussian one trough a
``gray" region where the differences between the two cases are not big.

\begin{figure}[t!]
\centering
\includegraphics[width=0.4\textwidth]{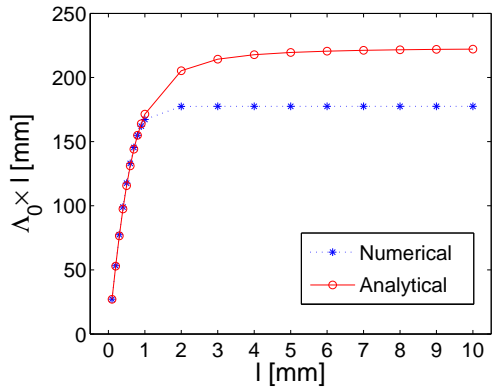}
\caption{(Color online) Numerical and analytical curves for the product $\Lambda_0 \times l$
versus the crystal length.}
\label{Lambda0 l vs l}
\end{figure}

Let us consider Eq. \eqref{LamGauss} in the limit of very large $l$.
In such case, since from Eq. \eqref{tau12_Gauss} $\tau_1\gg\tau_{\mathrm{p}}$%
, then $\Lambda_{0}$ asymptotically converges to:
\begin{equation}
\Lambda_0\simeq\pi^{1/4}\sqrt{20N_{\mathrm{p}}}\frac{\tau_{\mathrm{p}}}{%
|k^{\prime }_{\mathrm{p}}-k^{\prime }_{\mathrm{s}}|}\frac{1}{l}.
\end{equation}
This expression indicates that the product $\Lambda_0\times l$ is constant
for values of $l$ compatible with a non-Gaussian regime. In Figure \ref%
{Lambda0 l vs l}, then, we report the values of this product as a function
of the crystal thickness. For $l\lesssim1$mm the analytical solution, as
expected, is in good agreement with the numerical one, while for greater
thicknesses the discrepancy is significative. Also this result is expected,
since the analytical solution for the critical eigenvalue has not validity
when the condition \eqref{taupmin} is violated. On the other side, the fact
that also the analytical solution reaches asymptotically, for increasing $l$%
, a plateau suggests a $1/l$-like behavior of $\Lambda_0$. The existence of
such a plateau can be explained from the point of view of the evolution of
the pump and signal pulses in the time domain. As discussed in the previous
section, when the condition \eqref{taupmin} is violated, the walk-off
between pump and signal is bigger than the pump temporal width. As a
consequence, the exchange of energy between the two fields is disadvantaged
till a point where the threshold cannot change anymore even increasing the
crystal length. Since, from Eqs. \eqref{P0} and \eqref{Pthr}, $P_{\mathrm{thr%
}}\propto(\Lambda_0\times l)^{-2}$, for large values of $l$, then, also the
product $\Lambda_0\times l$ reach a constant value, thus explaining the
plateau in Fig. \ref{Lambda0 l vs l}.

In the same way, one can explain the discrepancy observed between
the two plateaux in Fig. \ref{Lambda0 l vs l}. As already discussed in case
C, in non-Gaussian configurations the pump bandwidth is larger than the
phase-matching one, thus not all the pump modes are phase matched and not
all of them contribute to the final value of the threshold. Therefore the
number of pump modes actually involved is smaller than the nominal value $N_{%
\mathrm{p}}$ that should, then, be corrected.

%The effective number of phase-matched modes can be evaluated from the ratio between the area beneath the eigenvector $L_0$ linked to $\Lambda_0$ and the pump spectral amplitude $\alpha$:
%
%\begin{equation}\label{Neff}
%N_{\text{eff}}=N_{p}\frac{\int_{-\infty}^{+\infty}\mathrm{d}\omega L_{0}\left(\omega\right)}{\int_{-\infty}^{+\infty}\mathrm{d}\omega \alpha\left(\omega\right)}
%\end{equation}
%Despite Eq. \eqref{Lam0} has been found in the Gaussian approximation limit, then, its prediction capability can still be extended to regions very far from this limit by considering the correction to the number of pump modes in order to keep in account only the phase-matched modes.

\begin{figure}[t!]
\centering
\includegraphics[width=0.4%
\textwidth]{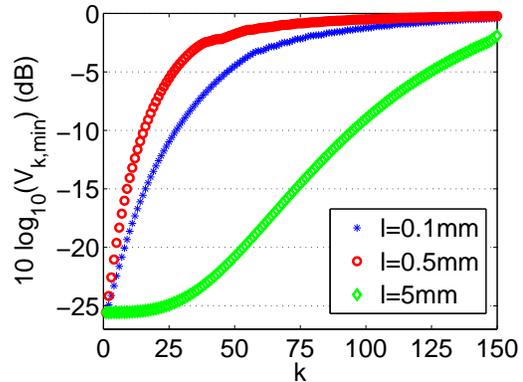}
\caption{(Color online) Comparison between cases A ($l=0.1$mm, middle curve), B ($l=0.5$mm,
upper curve) and C ($l=5$mm, lower curve). Noise variancese evaluated, at $%
\protect\omega=0$ and $r=0.9$, in the case of perfect mode matching of the
LO with the supermodes corresponding to the significative part of
eigenspectrum $0\leq k\leq150$. Among about $10^{5}$ supermodes only a
relatively small part is significative from a quantum point of view. In
particular, there are $45$, $23$ and $125$ supermodes, respectively, that
are squeezed more than $-5$ dB.}
\label{Vkmin_vs_l}
\end{figure}

From a quantum point of view, we have detailed in the previous sections the
noise properties of the supermodes connected to the first four highest $%
\Lambda_k$ (see Tables \ref{Vars09A}, \ref{Vars09B} and \ref{Vars09C}) and
we calculated the number of supermodes showing a squeezing better than $-5$
dB for getting a qualitative indication about the ``multimodicity" of the
system prepared in a specific experimental configuration. These results can
be appreciated in Fig. \ref{Vkmin_vs_l} where the noise variances of the
supermodes that satisfy this criterium have been traced for the three cases
previously discussed. The curve in the middle corresponds to case A ($l=0.1$%
mm), a Gaussian configuration. As the thickness of the crystal is increased
to $l=0.5$ mm (case B) the condition \eqref{taupmin} is violated but the
Gaussian approximation is still good. This means that, even if the value of $%
\Lambda_0$ is decreasing (because we are reducing the number of pump modes
that are phase-matched) the spectrum is still a geometric progression but
with a smaller, in absolute value, common ratio (see $\rho$ in Eq. %
\eqref{LamGauss}) thus causing an overall decrease of the spectrum with
respect to the case A. The consequence is a reduction of the number of
supermodes with a squeezing greater than $-5$ dB as it results from the upper
curve in Fig. \ref{Vkmin_vs_l} (red circles). Finally, when the crystal
length is further increased to $l=5$mm (case C), we pass to a completely
non-Gaussian configuration where the decrease of the critical eigenvalue $%
\Lambda_0$ (from $\sim271$ for $l=0.1$ mm to $\sim35$ for $l=5$ mm) causes a
significative deformation of the spectrum and a non null set of eigenvalues
flatten around $\Lambda_0$. In this case, since the degree of squeezing for
each supermode depends on the ratio $|\Lambda_k/\Lambda_0|$ (see Eq. %
\eqref{Squeeze}), the amount of squeezing is globally increased as the lower
curve (green diamonds) in Fig. \ref{Vkmin_vs_l} shows.

These results not only confirm that a SPOPO is a highly multi-mode device
but also show another important quality: the malleability for controlling
its ``multimodicity". We have seen, in fact, that one can just increase the
thickness of the nonlinear crystal in order to improve the number of
supermodes that play an important role from a quantum point of view.

In this paper we have presented a study of the eigenvalues and
eigenvectors of the matrix $\mathcal{L}$ in function of the crystal length $%
l $. However, since both the pump pulse width and the crystal length are
present in Eq. \eqref{taupmin}, notice that effects similar to those
discussed in this section can be observed by playing with $\tau_{\mathrm{p}}$
and keeping constant $l$. In fact the passage from a Gaussian to a
non-Gaussian configuration takes place when the pump bandwidth $N_{\mathrm{p}%
}$ becomes smaller than the phase-matching one $N_{1}$ and, clearly, this
can be obtained keeping fixed the latter and increasing $\tau_{\mathrm{p}}$
(see Fig. \ref{Legenda}). Eventually one could even fix $l$ and $\tau_{%
\mathrm{p}}$ and play with the group velocity mismatch by choosing different
types of nonlinearities, but, experimentally, this could result in a stiffer
malleability of the device.

\section{Conclusion}

In this paper, we have shown that both the dynamical and quantum noise
properties of the SPOPO depend on the spectrum of eigenvalues $\Lambda _{k}$
of the linear coupling matrix $\mathcal{L}$. We have studied in detail this
spectrum in various experimentally feasible configurations, using either an
analytical approach in some simple limit cases, or a numerical approach in
the general case. It turns out that among the roughly 100,000 eigenvalues,
99,900 or so are zero, and about a hundred are significantly different from
zero and contribute to the quantum dynamics of the system. SPOPOs are
therefore devices which produce simultaneously many highly squeezed vacuum
modes. This property can be used to improve the performances of metrological
methods using frequency combs, for example to perform ultra-accurate time
transfer between remote clocks beyond the shot noise limit \cite{Lamine}. In
addition, it is well known that if one mixes by one way or another different
squeezed modes, one gets strongly entangled states \cite{vanLoock}. This is also the case
here: we will show in a forthcoming publication that SPOPOs are indeed
likely to generate various pairs of strongly entangled supermodes, as well
as multipartite entangled states.

More generally, this paper is an example of the fact that, by using
appropriately chosen pump spectra and phase matching curves, one can reach
various eigenvalue spectra, and therefore tailor at will the quantum
properties of the light generated by the optical system, which can be useful
for example to generate interesting states for multidimensional quantum
information processing.

\section{Appendix : Quantum model for a singly resonant SPOPO}

We detail in this appendix the derivation of the Heisenberg equations in the
case of a singly resonant degenerate type I SPOPO, i.e. when the signal
field resonates inside the cavity, but not the pump field. We consider here
the case of the linear Fabry-Perot cavity. The treatment of this case is
more complex than the doubly resonant case usually considered in theoretical
approaches as the pump field cannot be quantized inside the cavity.

The signal field inside the nonlinear crystal, which extends from $z=-l/2$
to $z=+l/2$, is written as
\begin{equation}
\hat{E}_{\mathrm{s}}\left( z,t\right) =i\sum\limits_{q}\mathcal{E}_{\mathrm{s%
},q}\sin \left( k_{\mathrm{s},q}z^{\prime }\right) \hat{s}_{q}\left(
t\right) e^{-i\omega _{\mathrm{s},q}t}+\mathrm{H.c.},  \label{defEsB}
\end{equation}%
where $z^{\prime }=z+L/2$, and $\mathcal{E}_{\mathrm{s},q}=\sqrt{\frac{\hbar
\omega _{\mathrm{s},q}}{\varepsilon _{0}n\left( \omega _{\mathrm{s}%
,q}\right) A_{\mathrm{s}}L}}$. On the contrary, the pump field is not
affected by the cavity and hence it is given by a continuum of modes. We
shall use the common approach of quantizing the pump field in a line of
length $L_{\mathrm{p}}$ with periodic boundary conditions and, in the end of
the calculations, we will make $L_{\mathrm{p}}\rightarrow \infty $. We thus
write
\begin{eqnarray}
\hat{E}_{\mathrm{p}}\left( z,t\right)  &=&i\sum_{m}\mathcal{E}\left( \nu
_{m}\right)e^{-i\nu _{m}t}  \times   \notag  \label{defEp} \\
&\times &\left[ \hat{p}_{m}^{\left( +\right) }\left( t\right) e^{i\kappa
_{m}z}+\hat{p}_{m}^{\left( -\right) }\left( t\right) e^{-i\kappa _{m}z}%
\right]+ \nonumber\\
&+&\mathrm{H.c.},
\end{eqnarray}%
where the superscripts $\left( \pm \right) $ label the propagation
direction. $\mathcal{E}\left( \nu _{m}\right) =\sqrt{\frac{\hbar \nu _{m}}{%
2\varepsilon _{0}A_{\mathrm{p}}L_{\mathrm{p}}n\left( \nu _{m}\right) }}$ are
single photon field amplitudes. In the limit $L_{\mathrm{p}}\rightarrow
\infty $ the frequencies $\nu _{m}$ are given by $\nu _{m}=m\frac{2\pi c}{L_{%
\mathrm{p}}}$, $m\in \mathbb{N}$, and the wavenumbers $\kappa _{m}=\frac{%
n\left( \nu _{m}\right) \nu _{m}}{c}$, with $n\left( \nu _{m}\right) $ the
crystal refractive index. Note that we are writing the fields as a
superposition of plane waves, but the treatment is still approximately valid
for Gaussian beams as far as the thin crystal is placed at the (common) beam
waist of pump and signal, and the Rayleigh lengths are much longer than the
crystal length $l$. In such case $A_{\mathrm{f}}=\pi w_{\mathrm{f}}^{2}$
with $w_{\mathrm{f}}$ the corresponding beam radius.

The interaction Hamiltonian $\hat{H}_{\mathrm{I}}$ is calculated as usual as
in \eqref{HIdef}. Inserting the expressions of the second order nonlinear
electric polarizations, one obtains the following form of the interaction
Hamiltonian, in which $\chi$ is the second order nonlinear susceptibility
(whose dispersion is neglected):
\begin{eqnarray}  \label{HIlinear}
\hat{H}_{\mathrm{I}}\left(t\right)&=& i\frac{\varepsilon_{0}\chi lA_{\mathrm{%
I}}}{2} \sum_{m}\sum_{q}\sum_{j} \mathcal{E}\left(\nu_{j}\right)\mathcal{E}_{%
\mathrm{s},m}\mathcal{E}_{\mathrm{s},q}F_{m,q}^{j}\times  \notag \\
&\times& \hat{s}_{m}^{\dag}\left(t\right)\hat{s}_{q}^{\dag}\left(t\right) %
\left[\hat{p}_{j}^{\left(+\right)}\left(t\right)+\hat{p}_{j}^{\left(-%
\right)}\left(t\right)\right] \times  \notag \\
&\times& e^{i\left( \omega_{\mathrm{s},m}+\omega_{\mathrm{s}%
,q}-\nu_{j}\right)t} e^{-i\left(k_{\mathrm{s},m}+k_{\mathrm{s},q}\right)L/2}+
\notag \\
&+&\mathrm{H.c.,}
\end{eqnarray}
where we defined the phase-mismatch factor%
\begin{equation}
F_{m,q}^{j}=\frac{\sin\left[ \left( \kappa_{j}-k_{\mathrm{s},m}-k_{\mathrm{s}%
,q}\right) \frac{l}{2}\right] }{\left( \kappa_{j}-k_{\mathrm{s},m}-k_{%
\mathrm{s},q}\right) \frac{l}{2}}.
\end{equation}
In (\ref{HIlinear}) we dropped highly phase mismatched terms, as usual.

If we introduce new signal boson operators
\begin{equation}
\hat{s}_{m,\mathrm{new}}\left( t\right) =\hat{s}_{m}\left( t\right) e^{+ik_{%
\mathrm{s},m}L/2}
\end{equation}
the interaction Hamiltonian becomes as (\ref{HIlinear}) but without the
exponential $e^{-i\left( k_{\mathrm{s},m}+k_{\mathrm{s},q}\right) L/2}$. In
the following we will use the new operators but omit the superscript ``$%
\mathrm{new}$" for simplicity.

From the previous expression of the hamiltonian, one can derive following
Heisenberg equations governing the time evolution of the pump and signal
operators:
\begin{eqnarray}
\frac{d\hat{p}_{j}^{\left(\pm\right)}\left(t\right)}{dt}&=& -\frac{%
\varepsilon_{0}\chi lA_{\mathrm{I}}}{2\hbar} \sum_{m}\sum_{q}\mathcal{E}%
\left(\nu_{j}\right)\mathcal{E}_{\mathrm{s},m}\mathcal{E}_{\mathrm{s}%
,q}F_{m,q}^{j} \times  \notag \\
&\times& \hat{s}_{m}\left(t\right)\hat{s}_{q}\left(t\right)
e^{-i\left(\omega_{\mathrm{s},m}+\omega_{\mathrm{s},q}-\nu_{j}\right)t}, \\
\frac{d\hat{s}_{m}\left(t\right)}{dt}&=& \frac{\varepsilon_{0}\chi lA_{%
\mathrm{I}}}{\hbar} \sum_{q}\sum_{j} \mathcal{E}\left(\nu_{j}\right)\mathcal{%
E}_{\mathrm{s},m}\mathcal{E}_{\mathrm{s},q}F_{m,q}^{j}\times  \notag \\
&\times& \hat{s}_{q}^{\dag}\left(t\right) \left[\hat{p}_{j}^{\left(+\right)}%
\left(t\right)+\hat{p}_{j}^{\left(-\right)}\left(t\right)\right]\times
\notag \\
&\times& e^{i\left(\omega_{\mathrm{s},m}+\omega_{\mathrm{s}%
,q}-\nu_{j}\right)t} .  \label{s}
\end{eqnarray}

The integration of the pump equations yields
\begin{eqnarray}
\hat{p}_{j}^{\left(\pm\right)}\left(t\right)&=& \hat{p}_{\mathrm{free}%
,j}^{\left(\pm\right)} - \frac{\varepsilon_{0}\chi lA_{\mathrm{I}}}{2\hbar}
\sum_{m}\sum_{q} \mathcal{E}\left(\nu_{j}\right)\mathcal{E}_{\mathrm{s},m}%
\mathcal{E}_{\mathrm{s},q}F_{m,q}^{j} \times  \notag \\
&\times& \int\nolimits_{0}^{t}\mathrm{d}t^{\prime} \hat{s}%
_{m}\left(t^{\prime}\right)\hat{s}_{q}\left(t^{\prime}\right)
e^{-i\left(\omega_{\mathrm{s},m}+\omega_{\mathrm{s},q}-\nu_{j}\right)t^{%
\prime}},
\end{eqnarray}
where $\hat{p}_{\mathrm{free},j}^{\left(\pm\right)}=\hat{p}%
_{j}^{\left(\pm\right)}\left(0\right)$ is the source-free part of the pump
(the field impinging the nonlinear crystal).

Using the usual Wigner-Weisskopf approach, valid because the nonlinear
interaction is assumed to be instantaneous, and using the approximation
\begin{equation}
\frac{\sin\left(\frac{\Omega}{2}t\right)}{\frac{\Omega}{2}}\simeq
2\pi\delta\left(\Omega\right),
\end{equation}
we obtain a value of $\hat{p}_{j}^{\left( \pm\right) }\left( t\right)$ that
one can then insert in Eq. \eqref{s}:%
\begin{eqnarray}
\hat{p}_{j}^{\left(\pm\right)}\left(t\right)&=& \hat{p}_{\mathrm{free}%
,j}^{\left(\pm\right)}\left(t\right)+  \notag \\
&-& \frac{\pi\varepsilon_{0}\chi A_{\mathrm{I}}l}{\hbar} \sum_{m}\sum_{q}
\mathcal{E}\left(\nu_{j}\right)\mathcal{E}_{\mathrm{s},m}\mathcal{E}_{%
\mathrm{s},q} F_{m,q}^{j}\times  \notag \\
&\times& \hat{s}_{m}\left(t\right)\hat{s}_{q}\left(t\right)
\delta\left(\omega_{\mathrm{s},m}+\omega_{\mathrm{s},q}-\nu_{j}\right),
\label{psol}
\end{eqnarray}

We can now pass to the continuum limit. For that we define continuum pump
operators in the following way:
\begin{equation}
\hat{p}_{\mathrm{free}}^{\left( \pm\right) }\left( \nu_{j}\right) =\sqrt{%
\frac{L_{\mathrm{p}}}{2\pi c}}\hat{p}_{\mathrm{free},j}^{\left( \pm\right) },
\end{equation}
which verify
\begin{equation}
\left[ \hat{p}_{\mathrm{free}}^{\left( \pm\right) }\left( \nu_{j}\right) ,%
\left[ \hat{p}_{\mathrm{free}}^{\left( \pm\right) }\left( \nu_{k}\right) %
\right] ^{\dag}\right] =\frac{L_{\mathrm{p}}}{2\pi c}\delta_{j,k}\underset{%
L_{\mathrm{p}}\rightarrow\infty}{\longrightarrow}\delta\left( \nu
_{j}-\nu_{k}\right) .
\end{equation}

Transforming sums into integrals, we obtain the following equation for the
signal modes:
\begin{eqnarray}
\frac{d\hat{s}_{m}\left(t\right)}{dt}&=& \frac{A_{\mathrm{I}}}{\sqrt{A_{%
\mathrm{p}}}}l\chi\sqrt{\frac{\varepsilon_{0}}{4\pi\hbar c}} \sum_{q}
\mathcal{E}_{\mathrm{s},m}\mathcal{E}_{\mathrm{s},q} \hat{s}%
_{q}^{\dag}\left(t\right)I_{m,q}^{\left(1\right)}+  \notag \\
&-&\frac{A_{\mathrm{I}}^{2}}{A_{\mathrm{p}}}\left(l\chi\right)^{2}\frac{%
\varepsilon_{0}}{2\hbar c} \sum_{n}\sum_{r}\sum_{q} \mathcal{E}_{\mathrm{s}%
,m}\mathcal{E}_{\mathrm{s},q}\mathcal{E}_{\mathrm{s},n}\mathcal{E}_{\mathrm{s%
},r} \times  \notag \\
&\times& \hat{s}_{q}^{\dag}\left( t\right)\hat{s}_{n}\left(t\right)\hat{s}%
_{r}\left(t\right)I_{m,q}^{\left(2\right)},
\end{eqnarray}
where
\begin{eqnarray}
I_{m,q}^{\left(1\right)}&=& \int\mathrm{d}\nu \sqrt{\frac{\nu}{%
n\left(\nu\right)}} \frac{\sin\left[\left(k\left(\nu\right)-k_{\mathrm{s}%
,m}-k_{\mathrm{s},q}\right)\frac{l}{2}\right]} {\left(k\left(\nu\right)-k_{%
\mathrm{s},m}-k_{\mathrm{s},q}\right)\frac{l}{2}} \times  \notag \\
&\times& \left[ \hat{p}_{\mathrm{free}}^{\left(+\right)}\left(\nu\right)+%
\hat{p}_{\mathrm{free}}^{\left(-\right)}\left(\nu\right) \right]
e^{i\left(\omega_{\mathrm{s},m}+\omega_{\mathrm{s},q}-\nu\right)t}, \\
I_{m,q}^{\left(2\right)}&=& \int\mathrm{d}\nu\delta \left(\omega_{\mathrm{s}%
,n}+\omega_{\mathrm{s},r}-\nu\right)\frac{\nu}{n\left(\nu\right)} \times
\notag \\
&\times& \frac{\sin\left[\left(k\left(\nu\right)-k_{\mathrm{s},m}-k_{\mathrm{%
s},q}\right)\frac{l}{2}\right]} {\left(k\left( \nu\right)-k_{\mathrm{s}%
,m}-k_{\mathrm{s},q}\right)\frac{l}{2}} \times  \notag \\
&\times& \frac{\sin\left[\left(k\left(\nu\right)-k_{\mathrm{s},n}-k_{\mathrm{%
s},r}\right)\frac{l}{2}\right]} {\left(k\left(\nu\right)-k_{\mathrm{s},n}-k_{%
\mathrm{s},r}\right)\frac{l}{2}} \times  \notag \\
&\times& e^{i\left(\omega_{\mathrm{s},m}+\omega_{\mathrm{s},q}-\nu\right)t}.
\end{eqnarray}
In order to calculate the first integral, we write it as a sum over
frequency intervals of width $\Omega$ and centered at frequencies $\omega_{%
\mathrm{p},r}=2\omega_{0}+r\Omega$, $r\in\mathbb{Z}$. Thus $I_{1}$ becomes
\begin{eqnarray}
I_{m,q}^{\left(1\right)}&\simeq& e^{i\left(\omega_{\mathrm{s},m}+\omega_{%
\mathrm{s},q}\right)t} \sum_{r} \sqrt{\frac{\omega_{\mathrm{p},r}}{%
n\left(\omega_{\mathrm{p},r}\right)}} \times  \notag \\
&\times& \frac{\sin\left[\left(k_{\mathrm{p},r}-k_{\mathrm{s},m}-k_{\mathrm{s%
},q}\right)\frac{l}{2}\right]} {\left(k_{\mathrm{p},r}-k_{\mathrm{s},m}-k_{%
\mathrm{s},q}\right)\frac{l}{2}} \times  \notag \\
&\times& \int\nolimits_{r}\mathrm{d}\nu \left[\hat{p}_{\mathrm{free}%
}^{\left(+\right)}\left(\nu\right)+\hat{p}_{\mathrm{free}}^{\left(-\right)}%
\left(\nu\right)\right] e^{-i\nu t}
\end{eqnarray}
where $k_{\mathrm{p},r}=k\left( \omega_{\mathrm{p},r}\right) =k\left(
2\omega_{0}+r\Omega\right) $.

We now define new pump operators
\begin{equation}
\sqrt{2\pi}e^{-i\omega_{\mathrm{p},r}t}\hat{p}_{\mathrm{in},r}^{\left(
\pm\right) }\left( t\right) =\int\nolimits_{r}d\nu e^{-i\nu t}\hat {p}_{%
\mathrm{free}}^{\left( \pm\right) }\left( \nu\right) .  \label{defpin}
\end{equation}
which can be shown to verify the following property:
\begin{equation}
\left\langle \hat{p}_{\mathrm{in},r_{1}}^{\left( \pm\right) }\left(
t_{1}\right) \left[ \hat{p}_{\mathrm{in},r_{2}}^{\left( \pm\right) }\left(
t_{2}\right) \right] ^{\dag}\right\rangle \simeq\delta
_{r_{1},r_{2}}\delta\left( t_{1}-t_{2}\right).  \label{corr-p}
\end{equation}
$I_{m,q}^{\left(1\right)}$ now becomes
\begin{eqnarray}
I_{m,q}^{\left(1\right)}&\simeq& \sqrt{2\pi}\sum_{r} \sqrt{\frac{\omega_{%
\mathrm{p},r}}{n\left(\omega_{\mathrm{p},r}\right)}} \frac{\sin\left[\left(
k_{\mathrm{p},r}-k_{\mathrm{s},m}-k_{\mathrm{s},q}\right)\frac{l}{2}\right]%
} {\left(k_{\mathrm{p},r}-k_{\mathrm{s},m}-k_{\mathrm{s},q}\right)\frac{l}{2}%
} \times  \notag \\
&\times& e^{i\left(\omega_{\mathrm{s},m}+\omega_{\mathrm{s},q}-\omega_{%
\mathrm{p},r}\right)t} \left[ \hat{p}_{\mathrm{in},r}^{\left(+\right)}%
\left(t\right)+\hat{p}_{\mathrm{in},r}^{\left(-\right)}\left(t\right) \right]%
.
\end{eqnarray}

Retaining only slowly varying terms in the evolution, and including the
losses of the optical cavity at rate $\gamma_{\mathrm{s}}$, one finally
gets:
\begin{eqnarray}
\frac{d\hat{s}_{m}\left(t\right)}{dt}&=& -\gamma_{\mathrm{s}}\hat{s}%
_{m}\left(t\right) +\sqrt{2\gamma_{\mathrm{s}}}\hat{s}_{\mathrm{in}%
,m}\left(t\right)+  \notag \\
&+& g\sum_{q}f_{m,q}\hat{s}_{q}^{\dag}\left(t\right) \left[ \hat{p}_{\mathrm{%
in},m+q}^{\left(+\right)}\left(t\right)+\hat{p}_{\mathrm{in}%
,m+q}^{\left(-\right)}\left(t\right) \right]+  \notag \\
&-& g^{2}\sum_{n}\sum\limits_{q}f_{m,q}f_{n,m+q-n}\times  \notag \\
&\times& \hat{s}_{q}^{\dag}\left(t\right)\hat{s}_{n}\left(t\right)\hat{s}%
_{m+q-n}\left(t\right),  \label{dsmLan}
\end{eqnarray}
where the coupling constant $g$ is given by
\begin{equation}
g=\chi\frac{A_{\mathrm{I}}}{A_{\mathrm{s}}\sqrt{A_{\mathrm{p}}}} \frac{l}{L}%
\left(\frac{\omega_{0}}{n_{0}}\right) ^{3/2}\sqrt{\frac{\hbar}{%
\varepsilon_{0}c}}
\end{equation}
and $\hat{s}_{\mathrm{in},m}\left(t\right)$ corresponds to the field at
signal frequencies entering the cavity through the coupling mirror. When
that input is coherent or vacuum, the case we consider, those ``in"
operators verify the following correlation%
\begin{equation}
\left\langle \hat{s}_{\mathrm{in},m}\left( t\right) \hat{s}_{\mathrm{in}%
,m^{\prime}}^{\dag}\left( t^{\prime}\right) \right\rangle =\delta
_{m,m^{\prime}}\delta\left( t-t^{\prime}\right) ,
\end{equation}
and thus behave as $\hat{p}_{\mathrm{in},r}^{\left( \pm\right) }\left(
t\right) $ (see Eq(\ref{corr-p})).

Let us now consider the regime below the oscillation threshold : the signal
modes are almost not excited and the double sum in (\ref{dsmLan}) can be
neglected. Also, the pump ``in" fields can be approximated by their mean
values as their fluctuation part gives rise to smaller terms, which are
neglected for the same reasons as before. Hence, we have for a
unidirectional pumping :
\begin{equation}
<\hat{p}_{\mathrm{in},m}^{\left( -\right) }\left( t\right)> =0 \quad ; \quad
<\hat{p}_{\mathrm{in},m}^{\left( +\right) }\left( t\right)> =\sqrt {\frac{%
n_{0}A_{\mathrm{p}}P}{2\hbar\omega_{0}}}\alpha_{m},  \label{pinres}
\end{equation}
$P$ being the average power per unit area of the modelocked pump laser and $%
\sum _{m}\left\vert \alpha_{m}\right\vert ^{2}=1$. Eq. (\ref{pinres}) is
obtained by demanding that the pump field corresponding to the set $\left\{
\hat {p}_{\mathrm{in},m}^{\left( +\right) }\left( t\right) \right\} $ equals
the external pump field given by Eq. (\ref{Eext}) inside the crystal. The
linearized equations for the SPOPO below threshold finally become%
\begin{equation}
\begin{split}
\frac{d\hat{s}_{m}\left(t\right)}{dt}= &-\gamma_{\mathrm{s}}\hat{s}+\sqrt{%
2\gamma_{\mathrm{s}}}\hat{s}_{\mathrm{in},m}\left(t\right)+ \\
&+\gamma_{\mathrm{s}}\sigma\sum\limits_{q}f_{m,q}\alpha_{m+q}\hat{s}%
_{q}^{\dag}\left(t\right),  \label{dsdtsingle}
\end{split}%
\end{equation}
where
\begin{equation}
\sigma=\sqrt{\frac{P}{P_{0}}} \quad ; \quad P_{0}=\frac{%
\varepsilon_{0}c^{3}n_{0}^{2}T_{\mathrm{s}}^{2}}{32\left( \chi
l\omega_{0}\right) ^{2}}\left( \frac{A_{\mathrm{s}}}{A_{\mathrm{I}}}\right)
^{2},  \label{P0single}
\end{equation}

In conclusion, we have shown that the linearized equations (\ref{dsdtsingle}%
) formally coincide with those of a doubly resonant SPOPO (Eq.(\ref%
{dsdtdouble})), the only difference being the exact value of $P_{0}$.

\begin{acknowledgments}
Laboratoire Kastler Brossel, of the Ecole Normale Sup\'{e}rieure and the
Universit\'{e} Pierre et Marie Curie -- Paris6, is UMR8552 of the Centre
National de la Recherche Scientifique. This work was partially supported by
the French-Spanish Programme \textquotedblleft Partenariats Hubert
Curien"-\textquotedblleft Programa de Acciones Integradas" (Projet Picasso
13663PC - Acci\'{o}n Integrada HF2006-0018). GP, NT and CF
acknowledge the financial support of the Future and Emerging Technologies
(FET) programme within the Seventh Framework Programme for Research of the
European Commission, under the FET-Open grant agreement HIDEAS, number
FP7-ICT-221906. GJdeV acknowledges financial support by grant BEST/2007/161
of the Generalitat Valenciana and by Projects FIS2005-07931-C03-01 and
FIS2008-06024-C03-01 of the Spanish Ministerio de Educaci\'{o}n y Ciencia
and the European Union FEDER.
\end{acknowledgments}


\begin{thebibliography}{99}
\bibitem{lam0} Should $\Lambda_{0}<0$ then the null eigenvalue at threshold
would be $\lambda_{0}^{\left( -\right) }$ instead of $\lambda_{0}^{\left(
+\right) }$ and the following analysis should be accordingly modified.
\end{thebibliography}
\end{document}